\documentclass{article}

\usepackage[english]{babel}
\usepackage{natbib}
\usepackage[letterpaper,top=2cm,bottom=2cm,left=3cm,right=3cm,marginparwidth=1.75cm]{geometry}
\usepackage{amsmath}
\usepackage{amssymb}
\usepackage{graphicx}
\usepackage[colorlinks=true, allcolors=blue]{hyperref}
\usepackage{threeparttable} 
\usepackage[table]{xcolor}
\newcommand{\headrow}{\rowcolor{black!20}}
\usepackage{float}
\usepackage{caption}

\usepackage[inline]{enumitem}
\usepackage{dsfont}
\usepackage{xr}
\usepackage{tikz}
\usepackage{setspace}
\makeatletter
\newcommand\footnoteref[1]{\protected@xdef\@thefnmark{\ref{#1}}\@footnotemark}
\makeatother
\newcommand*\circled[1]{\tikz[baseline=(char.base)]{%
            \node[shape=circle,fill=white!20,draw,inner sep=2pt] (char) {#1};}}

\title{Modelling species distributions with deep learning to predict plant extinction risk and assess climate change impacts}
\author{Joaquim Estopinan, Pierre Bonnet, Maximilien Servajean, François Munoz, Alexis Joly}
\date{}

\begin{document}
\maketitle

\begin{abstract}
The post-2020 global biodiversity framework needs ambitious, research-based targets.
Estimating the accelerated extinction risk due to climate change is critical.
The International Union for Conservation of Nature (IUCN) measures the extinction risk of species.
Automatic methods have been developed to provide information on the IUCN status of under-assessed taxa.
However, these compensatory methods are based on current species characteristics, mainly geographical, which precludes their use in future projections.
Here, we evaluate a novel method for classifying the IUCN status of species benefiting from the generalisation power of species distribution models based on deep learning.
Our method matches state-of-the-art classification performance while relying on flexible SDM-based features that capture species' environmental preferences.
Cross-validation yields average accuracies of 0.61 for status classification and 0.78 for binary classification.
Climate change will reshape future species distributions.
Under the species-environment equilibrium hypothesis, SDM projections approximate plausible future outcomes.
Two extremes of species dispersal capacity are considered: unlimited or null.
The projected species distributions are translated into features feeding our IUCN classification method.
Finally, trends in threatened species are analysed over time and i) by continent and as a function of average ii) latitude or iii) altitude.
The proportion of threatened species is increasing globally, with critical rates in Africa, Asia and South America.
Furthermore, the proportion of threatened species is predicted to peak around the two Tropics, at the Equator, in the lowlands and at altitudes of 800-1,500 m.
\end{abstract}

\section{Introduction}
Failure to meet any of the 2020 Aichi Biodiversity Targets is a clear signal that transformative change is urgently needed \citep{ipbes_global_2019}.
The post-2020 global biodiversity framework must set ambitious targets with quantified, measurable objectives underpinned by research \citep{mace_aiming_2018}.
Biodiversity targets based on species extinction rates and measures of ecosystem services, for example, should be included \citep{rounsevell_biodiversity_2020, reyers_getting_2013}. Indeed, this could help to galvanise policy in a similar way to the 2°C maximum climate change target.
Finally, such goals are inherently interlinked and should be set together to allow parallel progress and avoid contradictions \citep{diaz_set_2020}.
Quantifying the acceleration of extinction risk due to climate change appears to be a top priority in this context \citep{mace_aiming_2018}.

\paragraph{Extinction risk and climate change.}
The increase in extinction risk due to climate change is an active area of research \citep{thomas_extinction_2004, malcolm_global_2006, carpenter_one-third_2008, maclean_recent_2011}.
A common practice is to use Species Distribution Models (SDMs) to first learn species' environmental preferences and then project the learned relationship into future climates \citep{guisan_predicting_2005}.
Species are then predicted to become extinct if their potential habitat is reduced below minimal thresholds.
Habitat loss rates can also be compared to the official extinction risk criteria of the IUCN \citep{mace_quantification_2008, moat_least_2019}.
While this approach is largely dominant, other ways of estimating extinction risk include using process-based models of physiology or demography, or relying on species-area relationships and expert opinion, as presented in this meta-study \citep{urban_accelerating_2015}.  Their literature review concludes that under Representative Concentration Pathway 8.5 (RCP~8.5), one in six species will be at risk of extinction due to climate change.
Based on socio-economic behaviours, RCPs are scenarios sampling the range of plausible greenhouse gas concentration until 2100 \citep{van_vuuren_representative_2011}.
RCP~8.5 is the scenario with the highest assumed fossil fuel use, but also the best match to our current emissions levels and stated policies \citep{schwalm_rcp85_2020}.
A wide range of climate model projections coexist for each RCP scenario.

\paragraph{The IUCN Red list of Threatened Species.}
The IUCN Red List of Threatened Species (RL) is the reference classification scheme for the extinction risk of species. The risk assessment is uniform for all living organisms (except micro-organisms, \citealt{bland2017guidelines}). It is a strength in terms of coherence and visibility.
The Red List is the basis for biodiversity indicators used in international agreements.
It allows monitoring of Parties' commitment to conservation.
Starting with \textit{least concern} (LC) and \textit{near threatened} (NT), the threat categories are ordered by increasing risk of extinction. Threatened species are either \textit{vulnerable} (VU), \textit{endangered} (EN) or \textit{critically endangered} (CR) before possibly becoming \textit{extinct} (EXT).

\paragraph{IUCN status classification.}
Manual assessment of extinction risk cannot keep up with the current levels of threats, so species are disappearing before they have been assessed or even discovered \citep{pimm_how_2015}.
As of 2022, only 15\% of the world’s known plant species have an IUCN status (\href{https://www.iucnredlist.org/about/barometer-of-life}{barometer of life}).
Research has responded to this massive concern by developing compensatory automated assessment methods \citep{stevart_third_2019}.
Such methods are designed to provide preliminary extinction risk categories and focus manual effort where it is urgently needed \citep{bachman_rapid_2020}.
Methods can either estimate IUCN variables to be compared with the official criteria thresholds (index-based methods) or directly learn a correspondence between species features and IUCN status (\textit{prediction-based} methods), \citet{zizka_automated_2020}. Classifiers such as the Random Forest (RF) algorithm and ensemble methods show high performance in differentiating threatened species within this active research area \citep{pelletier_predicting_2018, borgelt_more_2022}.
Species distribution modelling can inform IUCN assessments by estimating IUCN range variables (as recommended in the official guidelines, \textit{index-based} methods). Alternatively, SDMs and habitat modelling can be used to test new predictive methods of IUCN categories \citep{brooks_measuring_2019, breiner_including_2017}.
The latter authors concluded that SDM-based niche size estimates provide valuable complementary information to range size in assessing species extinction risk, but are not a good proxy for extent of occurrence (EOO) or area of occupancy (AOO). Other approaches therefore need to be explored.

\paragraph{Species distribution modelling.}
SDMs are statistical tools interlinking terrain observations with environmental variables \citep{elith_species_2009}.
Modelling species distributions involves approximations.
First, species are assumed to be in equilibrium with their environment when learning their potential niche and to remain so in the future \citep{araujo_equilibrium_2005}.
Second, data on dispersal capacity and direct species drivers are often lacking, leading to interval predictions and the use of proxy features.
Nevertheless, SDMs are already at the origin of numerous conservation successes \citep{guisan_predicting_2013, elith_statistical_2011}.
Moreover, the design of SDM is rapidly evolving to mitigate acknowledged limitations \citep{valavi_predictive_2022, rocchini_quixotic_2023}.
Building on its many successes in computer vision over the last decade, deep learning is now being applied to ecology to tackle complex tasks such as modelling species distribution \citep{lamba_deep_2019}.
The explosion of biodiversity data, resulting from both new techniques (citizen data science, remote sensing) and efforts to pool freely available data such as the GBIF (Global Biodiversity Information Facility), requires adapted modelling frameworks.
The deep learning (DL) community has been specifically designing such framework for years \citep{borowiec_deep_nodate}.
SDMs based on DL techniques allow to learn complex relationships between species and their environment \citep{botella2018deep, deneu_convolutional_2021}.
Finally, research to extend the predictive power of DL and to overcome challenges such as class imbalance, model interpretability, multi-modality fusion or label noise directly benefits species distribution modelling \citep{dosovitskiy_image_2020, cao_learning_2019, ryo_explainable_2021, benedetti_m3textfusion_2018, rolnick2017deep}.

\paragraph{Contributions.}
Our main contribution is a novel method for extracting species traits predictive of IUCN extinction risk status, which allows modelling and testing the impact of bioclimatic projections.
Based on a deep species distribution model, it achieves state-of-the-art classification performance while allowing to explore climate change scenarios and test how status distributions might evolve.
Thanks to this approach, we proceed by analysing how threatened species would be distributed across continents, latitudes and altitudes under RCP~8.5 and two extreme dispersal scenarios.
While the number of threatened species is projected to increase globally, some areas will be particularly affected:
Africa, Asia, South America, latitudes around both Tropics and the Equator, and finally the lowlands and intermediate altitudes between 800 and 1500m.\\

\section{Materials and methods}

\subsection{Method motivation}

Current extinction risk assessments of vascular plants mostly rely on their geographic distribution with IUCN's B criterion and the EOO/AOO measures.
As a result, the geolocation of observations takes up a central position in species subsequent IUCN category.
This affects automated methods that logically obtain the geographic information to be the best predictor of species extinction risk.
However, over-relying on species observation locations prevents exploring future scenarios to project species distribution shifts and forecast likely trends in species extinction risks.
Using an alternative IUCN criterion is often impossible because of data scarcity.
For instance, the application of criterion A measuring reductions of population size over time is hampered by the lack of repeated assessments and knowledge of species generation length.
A sensitive response is then to model future species distribution, retain most likely species presence according to a validated selection strategy and apply EOO/AOO thresholds. 
Yet, changes in SDM-predicted range size were found to be a poor surrogate for species EOO and AOO \citep{breiner_including_2017}.
Further research is therefore needed to determine if and how spatial predictions of SDMs can be used to adequately complement current IUCN assessments.

In response of this limitations, our goal is to extract species classification features with high generalisation capabilities in space and time.
To this end, we use a deep-SDM to reduce the dimension and capture critical environmental information that correlates with species observations.
The successive convolutional filters and pooling layers greatly reduced the input dimension, with the aim of capturing and preserving characteristic patterns.
The resulting $N$-dimensional space is hereafter called the \textit{feature space}.
A fundamental assumption in our method is that these features are not only informative about the species likely to be present conditionally on an observation, but also informative about the species' environmental and spatial niche.\\

\subsection{Data on the \textit{Orchidaceae} family}

\subsubsection{Species observations}
Orchid observations have been filtered from GBIF in \citep{zizka_automated_2020} thanks to the R package \textit{CoordinateCleaner}. \citep{zizka_coordinatecleaner_2019}.
The global dataset contains 999,258 occurrences of 14,129 species.
It is highly unbalanced: a few common species have thousands of observations, while most orchids are represented by a few opportunistic samples. The average number of occurrences per species is 70, while the median is 4. Only a quarter of the species have more than 13 occurrences.
The oldest occurrences date back to 1901, but three quarters were observed after 1982 (and half after 1997). Metadata on the orchid observation dataset are provided in the Appendix \ref{si:occ_distributions}.\\

\subsubsection{IUCN Red List status}
In July 2023, there are 1,970 orchid species assessed on the Red List, or 6.3\% of the estimated 31,000 species of this uniquely diverse family \citep{POWO}.
When we checked our orchid observations against the Red List as of December 2021, we found 889 species that had already been assessed. Figure~\ref{fig:boxplot}A shows the distribution of status. These species will be our reference for training our IUCN extinction risk classifier and evaluating its performance.
While the binary status distribution (threatened or not) is balanced, LC and EN species are largely dominant at the status level. Together, they represent more than two-thirds of the species assessed by IUCN.\\

\subsubsection{Predictive features}
Our predictive features include only global rasters at kilometre resolution. This makes the method easily transferable to other taxa. Large spatial contexts (64~x~64 km tensors) centred on each observation are provided to the model. The available predictors are bioclimatic variables from WorldClim2
variables, Soilgrids pedological variables, human footprint rasters, terrestrial ecoregions of the world and the observation location (see Appendix \ref{si:predictors} for details, input examples Fig.~\ref{si:input_exs} and full list of predictors Tab.~\ref{tab:features_table}).
No variable selection step was performed as deep convolutional neural networks do not overfit their predictive features under the right settings \citep{poggio_why_2017}.
Within the RCP~8.5 future scenario, the 19 WorldClim2 bioclimatic variables (averages from 1970-2000) are replaced by the corresponding projections from the \textit{EC-Earth3-Veg} climate model \citep{doscher_ec-earth3_2022}.
Four time periods are considered: 2021-2040, 2041-2060, 2061-2080 and 2081-2100.
More details can be found in the subsection \ref{ssec:projs}.\\

\subsection{Method definition}

\subsubsection{Deep species distribution modelling}
\label{sssec:deepSDM}
The first step consists in using a trained deep-SDM to encode the high-dimensional predictive data around each species occurrence into a reduced feature space (Figure~\ref{fig:scheme} \textbf{Step 1}).
The model used is an Inception v3 convolutional neural network \citep{szegedy_inception-v4_2016}.
The data set was divided into training/validation/test sets with spatial blocks of 0.025 degrees in the spherical coordinate system.
The 90/5/5\% block allocation was further stratified by region to optimise the diversity of the sets.
At the occurrence level this results in a set distribution of 902,174 / 46,290 / 50,794.
At the species level this leads to a distribution of 14,129 / 4,037 / 4,166.
Training was performed on two V100 GPUs from the Jean Zay supercomputer.
The model is trained with the LDAM loss, a modified cross-entropy function that gives more weight to rare species during training \citep{cao_learning_2019}.
Training on 70 epochs took 42 hours with a batch size of 128 and an initial learning rate of 0.01.
Model performance is evaluated every two epochs.
The final test set performance is reported for the best validation epoch.
Finally, the deep-SDM is retrained on the entire dataset for the best validation epoch prior to feature extraction.

Inception v3's multiple convolution and pooling layers allow the input information to be reduced to 2048 dimensions in the original version of the network. However, 2048 dimensions is still too many to perform classification on a few hundred samples.
To reduce the feature dimensionality, the Fully Connected (FC) layer with dimensions $(2048, \#labels)$ before the final softmax layer of the model was replaced by two layers $FC_1 = (2048, N), FC_2=(N, \#labels)$, creating a dimensional bottleneck.
ReLU activations are also appended after each FC.
Finally, the feature associated with an observation $\mathrm{o}$ is the $N$-dimensional test activation extracted after $FC_1$ and noted as $\mathrm{f}(\mathrm{o})$.

\paragraph{Deep-SDM validation.}
For a given observation, top-$k$ accuracy assesses whether the model returns the true label among the $k$ most likely species.
Success rates can then be calculated for all classes together (micro-average) or first by class and then averaged together (macro-average).
This means that the micro-average is more representative of performance on common species, whereas the macro-average is a better representation of performance on rare species.
Validation performance has plateaued since epoch 66, i.e. after the LDAM loss reweighting scheme in epoch 65.
The micro-average top-$30$ accuracy stabilises around $0.82$ and the macro-average top-$30$ accuracy stabilises around $0.42$, with the same performance on the test set.
The final deep-SDM is then retrained on the full dataset for 70 epochs.

\subsubsection{Dispersal scenarios}
\label{sssec:dispersal}
As a reminder, a fundamental assumption when inferring species distributions with an SDM is that species maintain the same environmental niche \citep{bakkenes_assessing_2002}.
The assumption that species have unlimited dispersal capacity leads to species niches shifting with climate change.
In practice, however, species have specific and limited dispersal capacities that prevent them from following climate change \citep{schloss_dispersal_2012}.
As data on plant dispersal capacity is extremely scarce, we worked with two extreme scenarios and assumed that the truth lies in between \citep{thomas_extinction_2004, urban_accelerating_2015}:
\begin{itemize}
    \item \textit{No dispersal.} Species can only be re-predicted by the deep-SDM at locations of true observations. By construction, this results in species potential presences (\textit{support points}) prior to SDM inference that can only be fewer in number than in the present.
    \item \textit{Unlimited dispersal.} Species can now be predicted at every location in the dataset (999,258 occurrences). In fact, the dataset is considered large enough to be used as an approximation of every possible location an orchid species could occupy.
\end{itemize}

In both dispersal scenarios, the relative probabilities of presence $\mathbb{P}(Y|X)$ returned by the SDM are thresholded to retain only the most likely species predictions (this is traduced by the indicator function in Equation~\ref{equ:wk}).
The threshold $\lambda$ was optimised as described in Appendix~\ref{si:modelCalib} to return precautionary species assemblages.
More specifically, the value was set on a calibration set to allow only a 3\% error in assessing whether the true label was retained in the assemblage - while keeping the threshold as high as possible. Then the model recall is optimised, which corresponds to the conformal prediction setting \citep{fontana_conformal_2023}.
As the threshold has been optimised to return likely overestimated but precautionary species assemblages, our estimates can be considered as lower bounds on species extinction risk.
Finally, species are classified as extinct if they are not predicted to occur at any point. Only the no dispersal scenario leads to this case.
A species that is predicted to become extinct in a given time period cannot return to any other status thereafter.

\subsubsection{Species niche features for IUCN classifcation}
\begin{figure}[htb!]
\centering
\includegraphics[width=\textwidth]{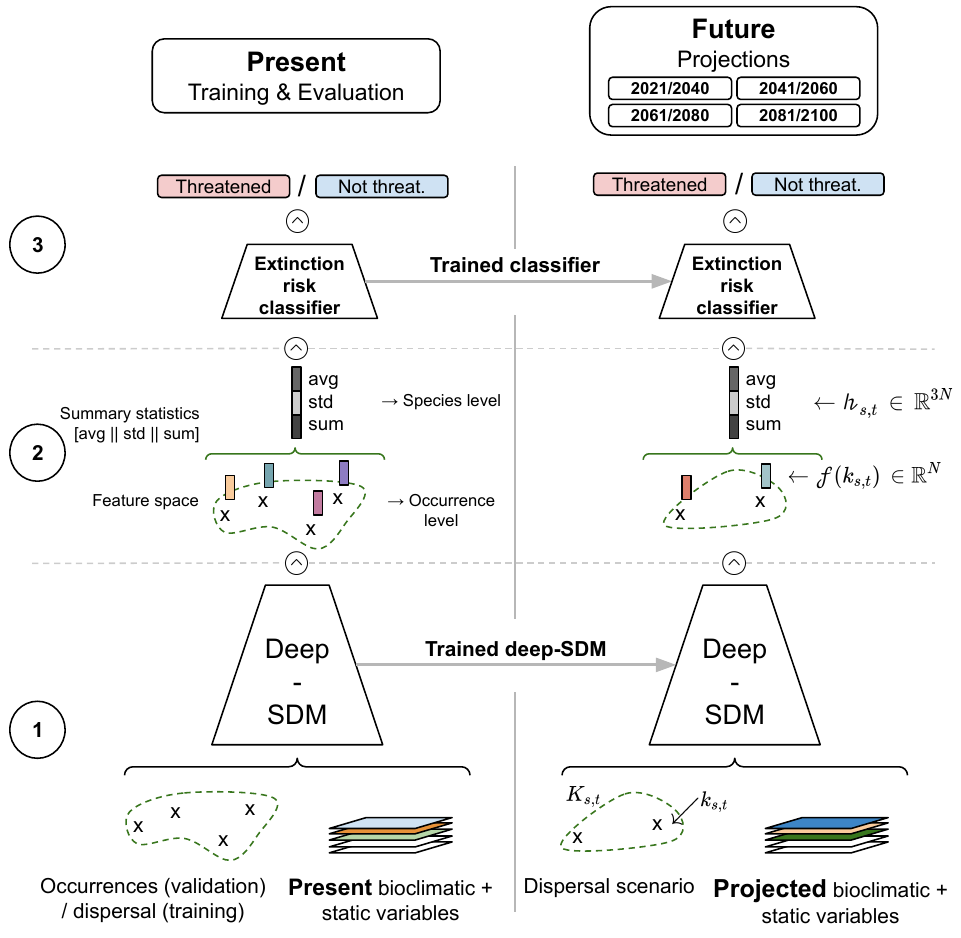}
\caption[Scheme of our extinction risk classifier fed with flexible SDM-based features]{Method scheme. The first column represents the current time frame (training and evaluation), while the right column represents a future scenario (projection).
\textbf{Step 1} consists in associating a given set of support points (true observations or dispersal scenario) with environmental covariates using deep-SDM inference. Species niches are indicated by dashed circles.
In \textbf{Step 2}, the predicted features are summarised by taking their mean, standard deviation and sum, and the result is concatenated. This operation allows the information that was at the point level to be condensed to the species level.
Finally, the \textbf{Step 3} is the mapping between the species summary feature and its conservation status. After training and validation in the present with real observations, the random forest classifier is trained within dispersal scenarios to ensure coherence with future projections. Classification can be either binary, as shown, or at the IUCN status level.}\label{fig:scheme}
\end{figure}

We need an information at the species level to feed an extinction risk classifier (Figure~\ref{fig:scheme} \textbf{Step 3}).
However, SDMs successively process environmental data at the level of an observation (Figure~\ref{fig:scheme} \textbf{Step 1}) before returning the most likely species in these conditions.
This is why we need to aggregate the information initially at observation level to species level.
To do so, we use summary statistics on vectors from the model's feature space (Figure~\ref{fig:scheme} \textbf{Step 2}) corresponding to a set of \textit{support points}: known points in the present or future potential presences.
In a future scenario, the choice of support points to aggregate features at the species level depends on species assumed dispersal capacities.

\noindent We now describe in more details the two first steps of our method for extracting species features. The item numbers are matching the steps of Figure~\ref{fig:scheme}.
\begin{enumerate}[label=\protect\circled{\arabic*}]
    \item \textbf{Identification of the support points of the species and deep-SDM inference.}
    At present, it is the species known sightings for model evaluation and status inference.
    In the future, the set of support points depends on the dispersal hypothesis made as detailed in section~\ref{sssec:dispersal}.
    Assuming no dispersal, the set of support points of a given species only includes the geolocations where the species has already been observed.
    Assuming conversely unlimited dispersal, the set of support points for a species is approximated by the 1M geolocations encompassed in our orchid observation dataset.
    Another possibility would have been to exploit a global regular grid as support points.
    However, we preferred to use the proxy of all the locations covered in our orchid dataset to limit computation resources.
    The deep-SDM was trained beforehand and evaluated with true occurrences and present covariates as described in section~\ref{sssec:deepSDM}.\\

    \item \textbf{Calculation of summary statistics on the resulting features.}
    Weighted mean, standard deviation and sum are computed over the features, along the $N$ dimensions.
    The final species feature is the concatenation of the three statistics above (dimension $3N$). Different summary statistics were evaluated. This concatenation led to the best results and will be further discussed in the final section.
    In addition, different weighting schemes for support points were considered when using a dispersal scenario.
    First, only points where the given species is predicted to be present with a minimum relative probability are retained, as explained in section~\ref{sssec:dispersal}.
    Second, among the retained points, their contribution is weighted according to the prediction rank of the target species.
    Indeed, a first implementation was to weight the contribution of the retained support points by the relative probability of presence of the target species.
    However, this strategy was not efficient to differentiate the support point contribution according to the most likely species.
    This is due to the deep-SDM calibration.
    The model was trained with presence-only observations and, as a result, the distribution of species relative probabilities has low dynamic.
    Therefore, for a given species, we preferred to weight the contribution of its support points with the inverse of its rank when ordering the most likely species conditional on the observation (see Equation~\ref{equ:wk}).\\
\end{enumerate}

\noindent Let $K_{s,t}$ be the set of support points of a species $s$ for the time period $t$ that depends on the dispersal scenario considered (see Section~\ref{sssec:dispersal}).
Based on the outputs of the deep-SDM at all points in $K_{s,t}$, we construct an aggregated feature vector $\mathrm{h}_{s,t}$ to be used as input of the final species status classifier:

\begin{equation}
    \mathrm{h}_{s,t} = \underset{k \in K_{s,t}}{\text{Stats}}[w_k.\mathrm{f}(k)]
\end{equation}
with:
\begin{itemize}[nosep]
    \item Stats = [avg $\mathbin\Vert$ std $\mathbin\Vert$ sum], $\mathbin\Vert$ being the concatenation operator
    \item $\mathrm{f}(k)$ the $N$-dimensional feature vector associated to the environmental covariates $x(k)$ through the deep-SDM.
    \item $w_k$ the weight attributed to support point $k$.
    The higher the weight, the more likely the species is present at this support point.
    $w_k$ expression depends on the species support points.
    We denote $\text{rank}[k_{s,t}]$ the position of species $s$ within the ordered list of the most likely species returned by the deep-SDM at the point $k$ and time period $t$. 
    \begin{equation}
    \label{equ:wk}
    w_k = \begin{cases}
      1               & \text{if true observations are used}\\
      \mathds{1}_{P [Y=s | X=x(k_{s,t})] \geq \lambda } \cdot \frac{1}{\text{rank}[k_{s,t}]} & \text{within a dispersal scenario}\\
    \end{cases}
\end{equation}
\end{itemize}

\subsubsection{Classifying extinction risk status from species niche features}

The final step of our method is to classify the extinction risk status of species from their aggregated feature vector $\mathrm{h}_{s,t}$:
\begin{enumerate}[label=\protect\circled{\arabic*}]
    \setcounter{enumi}{2}
\item \textbf{Extinction risk classification.} Therefore, a random forest classifier is trained using official IUCN status and the present occurrences of the species as support points.
In the present, it is then used in inference to determine the preliminary extinction risk status of unassessed species using their observations as support points.
To be coherent with the future dispersal scenario and for the calibration of the extinction risk classifier, the reference classifier used to predict future IUCN status is also retrained in the present with the same dispersal scenario (either null or unlimited).
Otherwise, a classifier that is i) trained with species features computed using true observations as support points, and ii) used to predict the future IUCN categories associated with species features aggregated from unlimited dispersal support points, will result in iii) severely underestimated extinction risk levels.
Finally, two levels of classification are considered: one at the binary level (threatened or not, as shown in the scheme) and another at the IUCN status level.
\end{enumerate}

\noindent Other classifiers were also considered before selecting the random forest: a shallow multilayer perceptron, a multinomial log-linear regression and linear classifiers with Stochastic Gradient Descent (SGD) training.
Performance was evaluated using a 10-fold cross-validation strategy on the 889 species assessed by IUCN.
The best classifier, the random forest, was then compared with the state-of-the-art \textit{IUCNN} extinction risk classifier \citep{zizka_iucnn_2021} at both binary and status levels (see model valiation Section~\ref{ssec:mod_validation}).
Once the relationship between species features and extinction risk status has been learned for the 889 IUCN-assessed species, risk status can be predicted for the remaining 13,240 species in the dataset.

\subsubsection{Projections within the RCP~8.5 scenario}
\label{ssec:projs}

\paragraph{Climate model choice.}
The choice of climate model was made using the GCMeval tool \citep{parding_gcmeval_2020}. The focus region was set to global and the skill assessment weights were left at their default values (equal importance given to temperature and precipitation, idem for all seasons and skill scores).
Finally, the greenhouse gas concentration scenario was set to RCP~8.5 and only climate models whose projections were available for download at i) 30 seconds spatial resolution, ii) for all time periods and iii) for the 19 bioclimatic variables on \href{https://www.worldclim.org/data/cmip6/cmip6_clim30s.html}{https://www.worldclim.org/} were considered.
This led to the adoption of the EC-Earth3-Veg system model projections \citep{doscher_ec-earth3_2022}.

\paragraph{Levels of analysis over time: Continents, latitude and altitude.}

In addition to the temporal dimension, the predicted status distributions are crossed with three other variables. One is categorical in the inhabited continents and two are continuous in the species mean latitude and mean altitude.
Inhabited continents were obtained by spatially intersecting occurrences with WGSRPD level 3 zones. A given species may span several continents.
Mean latitude was calculated directly from the observation coordinates.
Finally, elevation values were obtained from a global raster with 15 arc-second spatial resolution downloaded by tile at \href{http://www.viewfinderpanoramas.org/dem3.html}{http://www.viewfinderpanoramas.org/} and processed using GDAL command lines.
The predictions of extinction risk are identical to the three facets of the analysis. The only difference is in the presentation of the results.
In practice, it is the variable used to group species that varies from one representation to another.\\

\subsection{Model validation}
\label{ssec:mod_validation}

\begin{figure}[htb!]
    \centering
    \begin{minipage}{.45\textwidth}
        \centering
        \hspace{0.5cm} \textbf{A}\\
        \includegraphics[width=\linewidth]{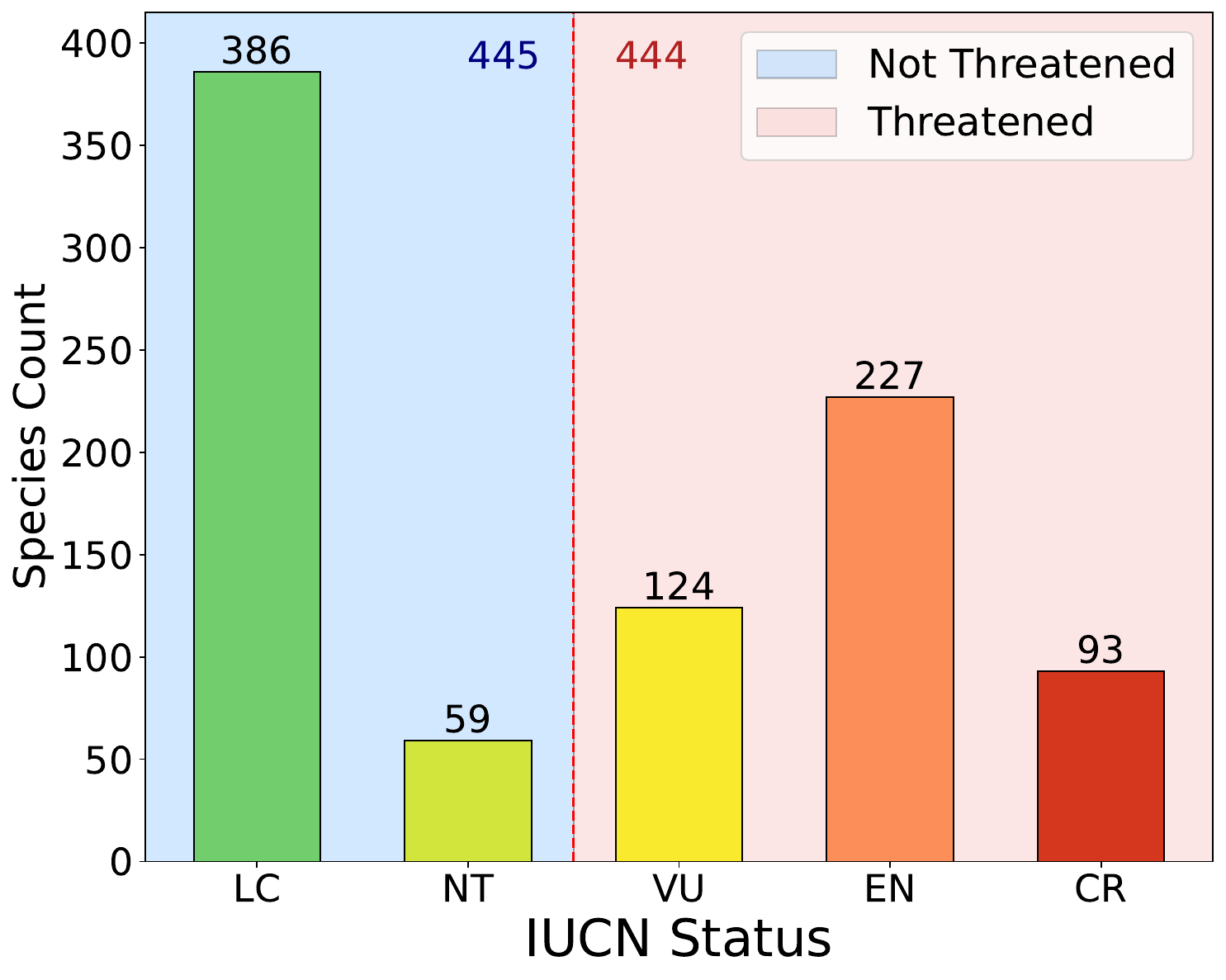}
    \end{minipage}%
    \begin{minipage}{0.55\textwidth}
        \centering
        \hspace{1cm} \textbf{B}\\
        \includegraphics[width=\linewidth]{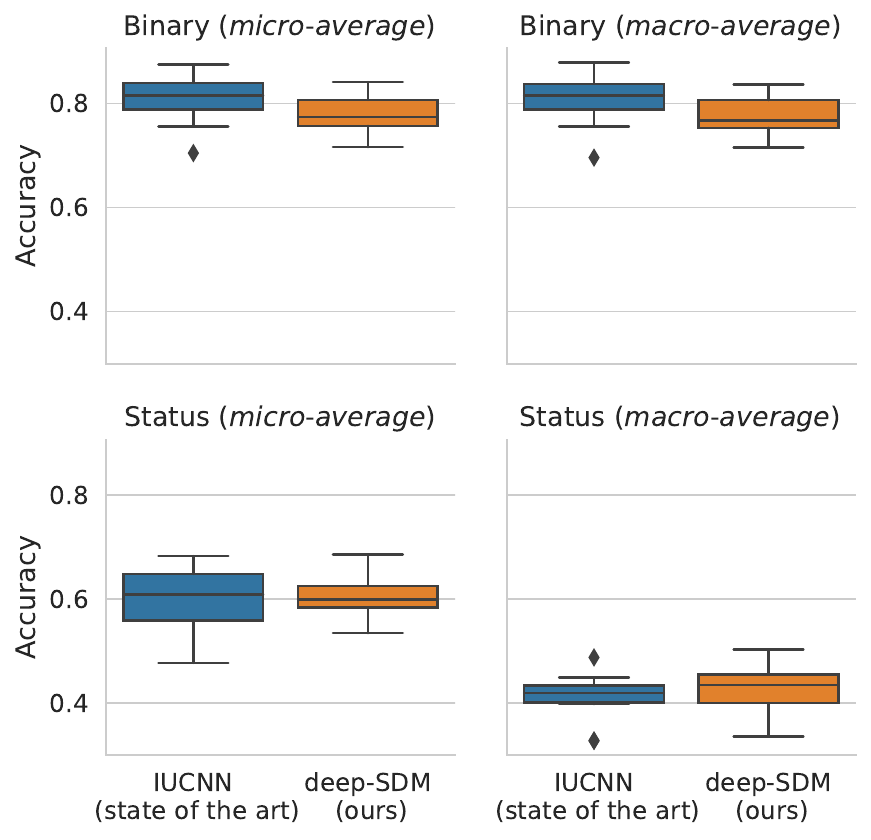}
    \end{minipage}
    \caption[Dataset IUCN status distribution and classification performance]{
    \textbf{A.} Distribution of the 889 IUCN extinction risk status from our dataset.
    \textbf{B.} Classification performance comparison between the state-of-the-art \textit{IUCNN} method and ours \citep{zizka_iucnn_2021}. The 10-fold cross-validation results in an accuracy distribution represented by boxplots. The first row shows the binary classification and the second row the status classification. Micro- and macro-average accuracies are identical, as the binary status distribution is balanced. The IUCNN method achieves an average accuracy of 0.81 for binary classification and our deep SDM method achieves 0.78. However, for status classification, our method gives a micro-average accuracy of 0.61 and a macro-average accuracy of 0.43 and the IUCNN method 0.60 and 0.41 respectively.}
    \label{fig:boxplot}
\end{figure}

Classification performance and comparison with the state-of-the-art IUCNN method are shown in Figure~\ref{fig:boxplot}B \citep{zizka_iucnn_2021}.
While each method has a slight advantage in either binary or status classification, their performances are close enough to consider the deep SDM-based method as competitive with the state-of-the-art method.
In addition, deep-SDM confusion matrices for the two classification levels are provided Fig.~\ref{si:cf_mat}.\\

\section{Results}
\label{sec:results_classif}
We analyse the global dynamics of orchid IUCN status distribution over time and in function of three aspects: continents, latitude and altitude.
In the figures below:
\begin{itemize}[nosep]
    \item[i)] the results from the two dispersal scenarios are averaged to provide synthetic trends (Figure~\ref{fig:conts} represents their difference with error bars)
    \item[ii)] all species are considered to be aiming at broad conclusions at the family level (i.e. both those already assessed by the IUCN and those that have predicted extinction risk categories even in the present)
    \item[iii)] binary extinction risk levels are reported as their prediction is more robust.
\end{itemize}
The same analysis, restricted to species assessed by the IUCN, is presented in Appendix~\ref{si:iucn-assessed_only}.
Comparing these gives a sense of the generalising power of our approach.\\

\subsection{Continents}

\begin{figure}[htb!]
    \centering
    \includegraphics[width=\linewidth]{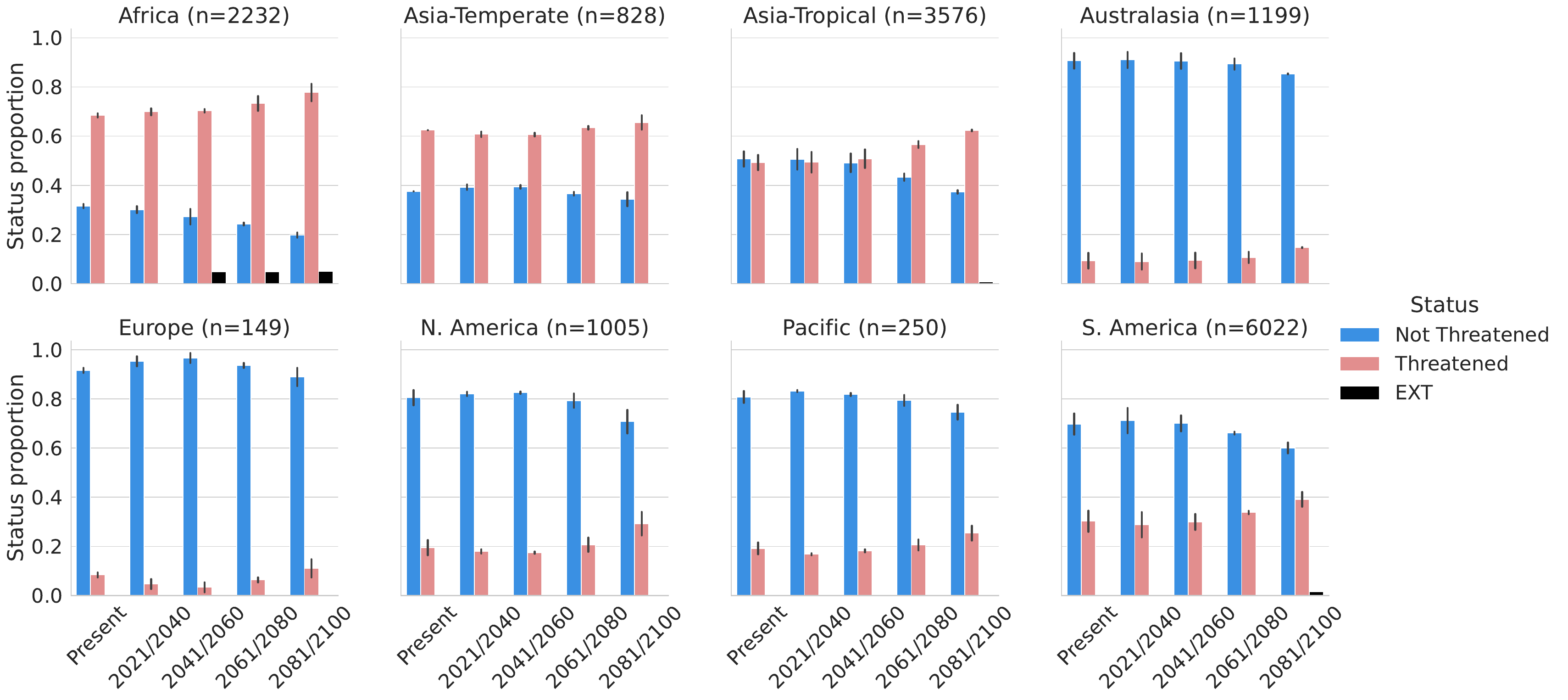}
    \caption[Status proportions per continent and time period]{Binary status proportions per continent and time period. All species are included and their number per continent is given in the subtitles. Error bars account for differences between the two dispersal scenarios.}
    \label{fig:conts}
\end{figure}

The global trend in Figure \ref{fig:conts} shows an increasing proportion of threatened species across all continents.
Individual patterns also appear:
\begin{itemize}
    \item Africa and Asia-Temperate are the only two continents with a current majority of threatened orchid species.
    \item Five percent of African species are predicted to become extinct between 2041 and 2060 (11\% of IUCN assessed species).
    \item In tropical Asia, threatened species become the majority by mid-century, reaching 60\% by the end of the century.
    \item Several continents - notably Europe and North America - see their proportion of threatened species decline in the first half of the century, before recovering to overtake current levels.
\item Although South America may appear relatively spared, its increase in the proportion of threatened species is significant and covers six thousand species.
\end{itemize}

On a global scale, the number of threatened species is expected to increase by an average of one third by the end of the century (14-40\% rise depending on dispersal scenario).
All status predictions are provided in the supplementary file \textit{ALL\_species\_status.csv}.
A smaller increase in threatened species is predicted for IUCN-assessed species, see Fig.~\ref{si:global_threat}.
Unassessed species could therefore be expected to be at even greater risk of extinction than those already assessed.

We also predict that 234 species will become extinct, of which 42 are IUCN-listed: 111 in Africa, 96 in South America, 26 in tropical Asia and one in temperate Asia.
The list is also given as supplementary file \textit{EXT\_species.csv}.
A small number of species have been identified in GBIF as having very few and old occurrences.
These two .csv files are described in Appendix \ref{si:status_pred}.

\subsection{Latitude}

\begin{figure}[htb!]
    \centering
    \includegraphics[width=\linewidth]{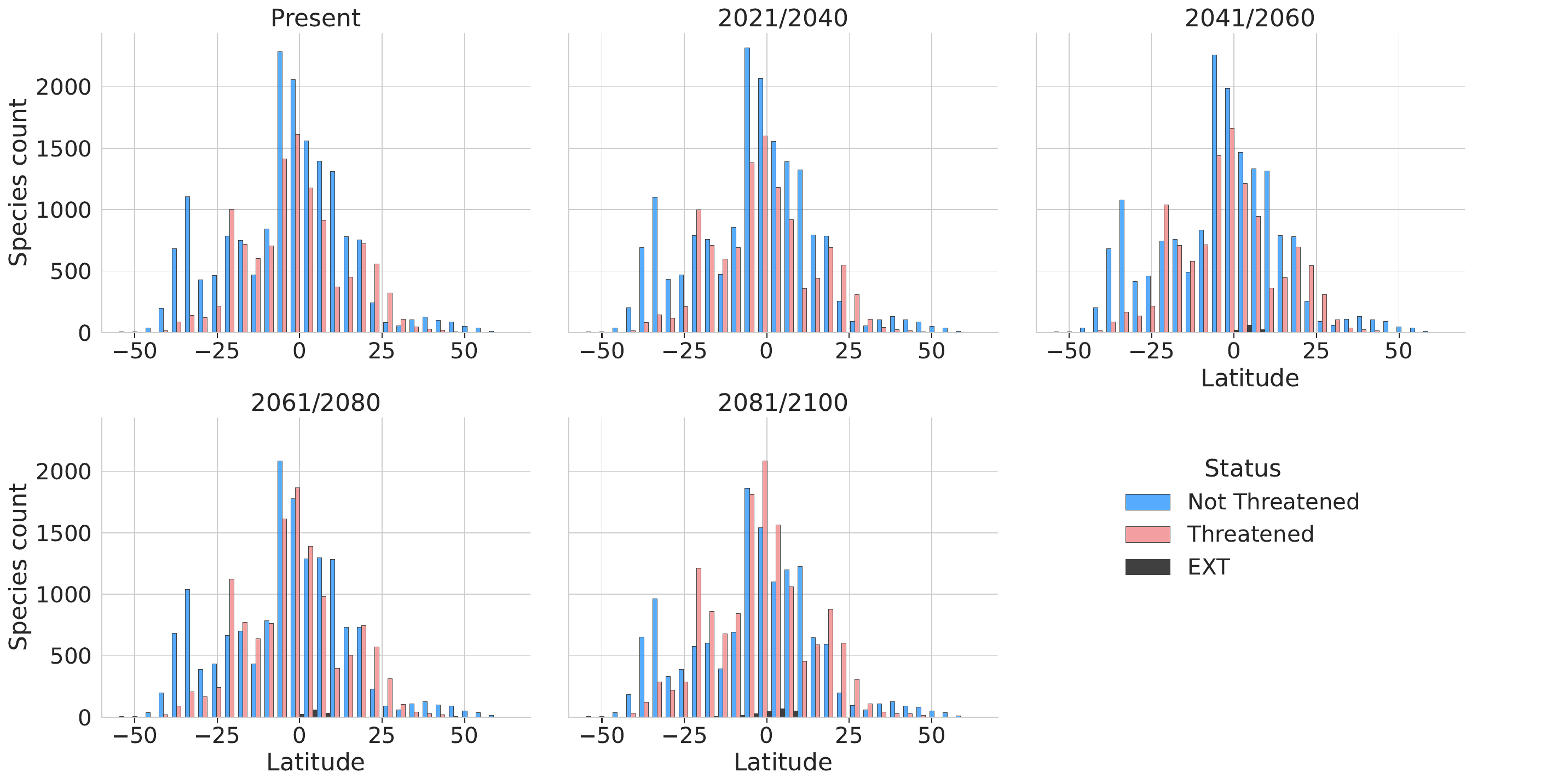}
    \caption[Status predictions in function of species average latitude]{Species count histograms as a function of average latitude and time period. All species are included and colours indicate the binary extinction risk status. Bins cover four degrees of latitude.}
    \label{fig:lat}
\end{figure}

In Figure \ref{fig:lat}, the number of threatened species clearly peaks around the equator and in the tropics, i.e. at latitudes with high species diversity. Species going extinct are also found around the equator. Both high and low latitudes are dominated by non-threatened species.\\

\subsection{Altitude}

\begin{figure}[htb!]
    \centering
    \includegraphics[width=\linewidth]{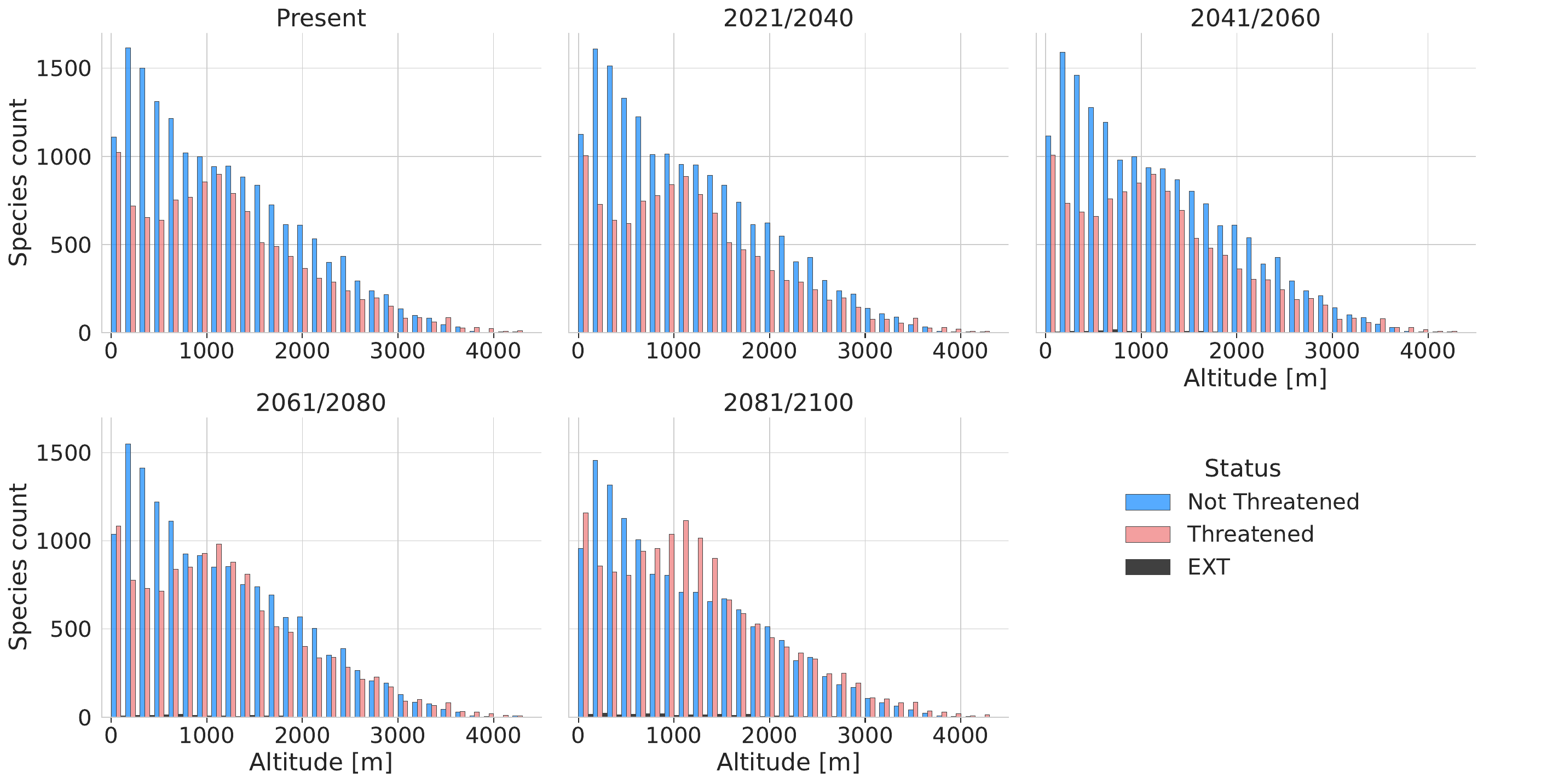}
    \caption[Status predictions in function of species average elevation]{Species count histograms as a function of their average altitude and over time. All species are included and colours indicate the binary extinction risk status. Bins cover 150 metre elevation ranges.}
    \label{fig:alt}
\end{figure}

Figure \ref{fig:alt} crosses the number of classified species over time with their average altitude.
The profile of threatened species along the altitudinal gradient appears to be stable over time. It shows a peak at low elevations and a concentration of threatened species in the 800-1500 m range.
However, the number of threatened species increases at all altitudes.
They outnumber non-threatened species in their peaks and also at very high elevations, about 2500~m.\\

\section{Discussion}

\subsection{Interpretations}
We now propose ecological interpretations that account for the different patterns observed.
The relative decline of threatened species by mid-century - particularly pronounced in Europe and North America - is rooted in the unlimited dispersal scenario.
Indeed, in these projections, the support points for species training and then potential niches can be largely overestimated, leading to lower rates of threatened species.
However, this phenomenon is mitigated in the second half of the century, where we assume that bioclimatic change significantly limits species' potential niches.
The large proportion of threatened species currently predicted for Africa and Asia-Temperate could be partly explained by a different accounting of LC species. Further analysis at this level is needed to test this possibility.
This seems all the more relevant when considering the proportion of threatened species restricted to IUCN-assessed species in Fig.~\ref{fig:conts_iucn}.

On average, considering both dispersal scenarios together, the number of threatened species is expected to increase by a third by 2100 with our method (14-40\% rise interval, see Figure~\ref{si:global_threat}).
This trend may seem small compared to other studies on the extinction risk due to climate change.
For example, \citet{urban_accelerating_2015} concluded that under current emissions trajectories, up to one in six species could be threatened with extinction by climate change.
However, the global trends we have expressed are relative, i.e. the increase in threatened species relative to current levels.
In absolute terms, we predict an 11\% increase in the number of all threatened species worldwide (6\% if only IUCN-assessed species are considered).
This is closer to the 16\% absolute increase from \citet{urban_accelerating_2015}.
Finally, as observed in the analysis of trends by continent, the impact of climate change on species loss is predicted to be greater in the second half of the century than in the first.
Similarly, \citet{van2006future} predicts that the impact of climate change will become increasingly important after 2050.

Species that are already IUCN-assessed are predicted to experience a smaller increase in their threatened share than species that are not. We formulate two hypotheses for this pattern.
First, plant assessment targets include a majority of species expected to be threatened \citep{bachman_progress_2019}, resulting in an overall proportion of threatened plant species of 48\%. In comparison, the global estimate is only ~21\% \citep{brummitt_green_2015}. This assessment bias therefore contributes to explaining the more stable proportion of threatened species among IUCN-assessed species.
Second, since the classifiers were trained on the IUCN-assessed species with current bioclimatic values, we can expect some overfitting on these species.
Even if bioclimatic variables change in the future, SDMs and consequently the extinction risk classifiers are also provided with static variables, which could indeed explain a higher tendency to re-predict status quo for these species.

The overall latitudinal species distribution follows, as expected, the latitudinal gradient of biodiversity \citep{willig_latitudinal_2003}.
However, we found no satisfactory hypothesis to explain the trident shape of the distribution of threatened species along the latitudinal gradient.
This shape is also present when only considering IUCN-assessed species, but with the highest number of threatened species peaking around the -20° parallel.

In the case of lowland species at risk, a direct assumption is that land-use change and high exposure to anthropogenic threats are at play.
The hump-shaped pattern of threatened species corresponds to the diversity peak known to occur at intermediate altitudes \citep{whittaker_vegetation_1960}.
Indeed, these altitudes are known to be rich transition zones between different habitats and with specific interactions between temperature and water gradients \citep{zhao_altitudinal_2005}.
However, this does not explain why only threatened species follow this pattern.
There may be several reasons.
First, speciation rates and endemicity also peak at intermediate altitudes, thanks to an optimal combination of area and isolation for the persistence and divergence of native species according to \citet{lomolino_elevation_2001}.
This would result in species with few individuals and/or restricted geographical ranges that are likely to be threatened.
Secondly, there is almost no more primary forest at low elevations, in contrast to intermediate elevations. These forests may have been orchid refuges in the past, but may now be increasingly threatened by extensive land change and climate change.
Finally, this irregular prediction of threatened species at mid-altitude could hide strong spatial discrepancies, for instance a particularly high increase in the Andes but not in the Alps.

Preliminary results show that terrestrial Human Footprint appears to correlate with the predicted proportion of threatened species, see Appendix~\ref{si:hfp_section}.
This is consistent with previous work showing a strong correlation between human footprint and species extinction risk (\cite{di_marco_changes_2018}.
This study even found anthropogenic pressure to be more predictive of extinction risk than environmental or life-history variables.

\subsection{Limitations}
Our method has limitations that we acknowledge below.
As previously mentioned, the threshold for species re-prediction may be too permissive.
In both dispersal scenarios, this may lead to an overestimation of geographic support for learning species features. A more restrictive choice could have the effect of increasing the number of species predictions at risk - further work is indeed necessary.
Similarly, the weighting scheme used to compute species features from activations needs more attention. Classification performance was indeed found to be sensitive to changes at this level and validation would allow the most appropriate scheme to be set.
Finally, SDMs rely on correlation patterns between observations and features, whereas explicit causal links would greatly facilitate the interpretation of results.

Furthermore, our species features are based on points projected in time and space for the dispersal scenario.
In the worst case, future bioclimatic conditions combined with static variables create previously unseen contexts leading to out-of-domain model inference.
In a conservative scenario, static variables lead to automatic re-prediction of true labels plus likely species, regardless of bioclimatic conditions. This results in constant or underestimated extinction risk depending on the dispersal scenario.
Finally, in the targeted scenario, bioclimatic projections combined with static variables shift species contexts in a large enough and structured representation domain, leading to coherent inferences.
Thanks to the size of our dataset and previous interpretability study on the generalisation power of deep-SDMs, we defend that the behaviour of the model does indeed tend towards this third scenario.

At the level of status classification, performance drops for NT and VU statuses, i.e. for transition categories (see confusion matrices Fig.~\ref{si:cf_mat}). While these statuses have relatively low training support (59 and 124 species respectively), we believe that this is largely due to the rather ambiguous definition of the NT category:
\vspace{-0.25cm}
\begin{quote}{\cite{bland2017guidelines}}
    \textit{“A taxon is Near Threatened when it has been evaluated against the criteria but does not qualify for Critically Endangered, Endangered or Vulnerable now, but is close to qualifying for or is likely to qualify for a threatened category in the near future.”}
\end{quote}
\vspace{-0.25cm}
As the reference delimitation between NT and VU species is potentially confused, their respective learning of their environemental representation is affected.
Performance optimisation eventually leads to their abandonment in favour of the more clearly defined LC/EN/CR class predictions.

\subsection{Perspectives}
The perspectives opened up by this study are many.
The SDM-based species features extracted before the final softmax layer of the model were shown to be predictive of extinction risk status.
This feature space close to the SDM output is meant to be linearly separable by the different classes.
In contrast, activations closer to the SDM input might be less informative about the classes, but more representative of the predictive features. Extracting species features earlier may be advantageous given that the activations flow on a predictors-class information gradient across the model layers.

One of the strengths of our model is its scalability to thousands of species.
However, it is not adapted to follow the IUCN species-specific guidelines when using an SDM to estimate the different criteria \citep{bland2017guidelines}.
A direct exchange with the conservation community on their specific needs would certainly help to guide future development efforts.
Another natural area for improvement is to focus on a few well-documented species and subject our method to the IUCN guidelines.
Furthermore, it is appealing to test our method on another suitable taxon to evaluate its taxonomic generalisation power.

Finally, including additional information that predicts the likely presence and extinction risk of species are other promising directions: species traits \citep{bourhis_explainable_2023}, ecosystem functional attributes \citep{arenas-castro_assessing_2018} and other threat/habitat rasters such as urban expansion forecasts, forest cover predictions or protected areas \citep{borgelt_more_2022, vieilledent_spatial_2022}.
Indeed, the inclusion of climate change alone inevitably leads to an underestimation of future threats to biodiversity \citep{brook_integrating_2009}.\\

\subsection{Conclusion}
The prediction of species extinction risk from SDM-based features achieves state-of-the-art performance while being flexible enough to allow testing climate change scenarios.
This means that the valuable information provided by the predictors has been successfully encoded by the SDM based on deep learning.
Future projections of orchid extinction risk averaged over two dispersal scenarios provide biodiversity trends to support global conservation targets \citep{nicholson2019scenarios}.
Indeed, this classification framework allows to investigate the impact of climate change on the distribution of species extinction risk.
While the proportion of threatened species is increasing globally, analysis by continent, latitude or altitude reveals specific and escalating patterns.\\

\section*{Acknowledgements}
The research described in this paper was funded by the European Commission via the GUARDEN and MAMBO projects, which have received funding from the European Union’s Horizon Europe research and innovation programme under grant agreements 101060693 and 101060639.
The opinions expressed in this work are those of the authors and are not necessarily those of the GUARDEN or MAMBO partners or the European Commission.
The INRIA exploratory action CACTUS fund also supported this work.
This work was granted access to the HPC resources of IDRIS under the allocation 20XX-AD011013648 made by GENCI.
Finally, we warmly thank Alexander Zizka for providing us with the filtered set of orchid occurrences.

\section*{Data availability statement}
Code and data are available at \href{https://github.com/estopinj/IUCN_classification}{github.com/estopinj/IUCN\_classification}.

\bibliographystyle{apalike}
\bibliography{sample}
\clearpage

\appendix
\section*{Supplementary information}

\renewcommand{\thefigure}{S\arabic{figure}}
\renewcommand{\thetable}{S\arabic{table}}
\renewcommand{\theequation}{s\arabic{equation}}
\setcounter{figure}{0}
\setcounter{table}{0}
\setcounter{equation}{0}

\section{Orchid dataset distributions}
\label{si:occ_distributions}
\begin{figure}[H]
    \vspace{-0.5cm}
    \hspace{3.5cm} \textbf{(a)} \hspace{6.cm} \textbf{(b)}
    \vspace{-0.3cm}
    \begin{center}
    \includegraphics[trim={0.2cm 0 0.3cm 0}, clip, width=6.5cm]{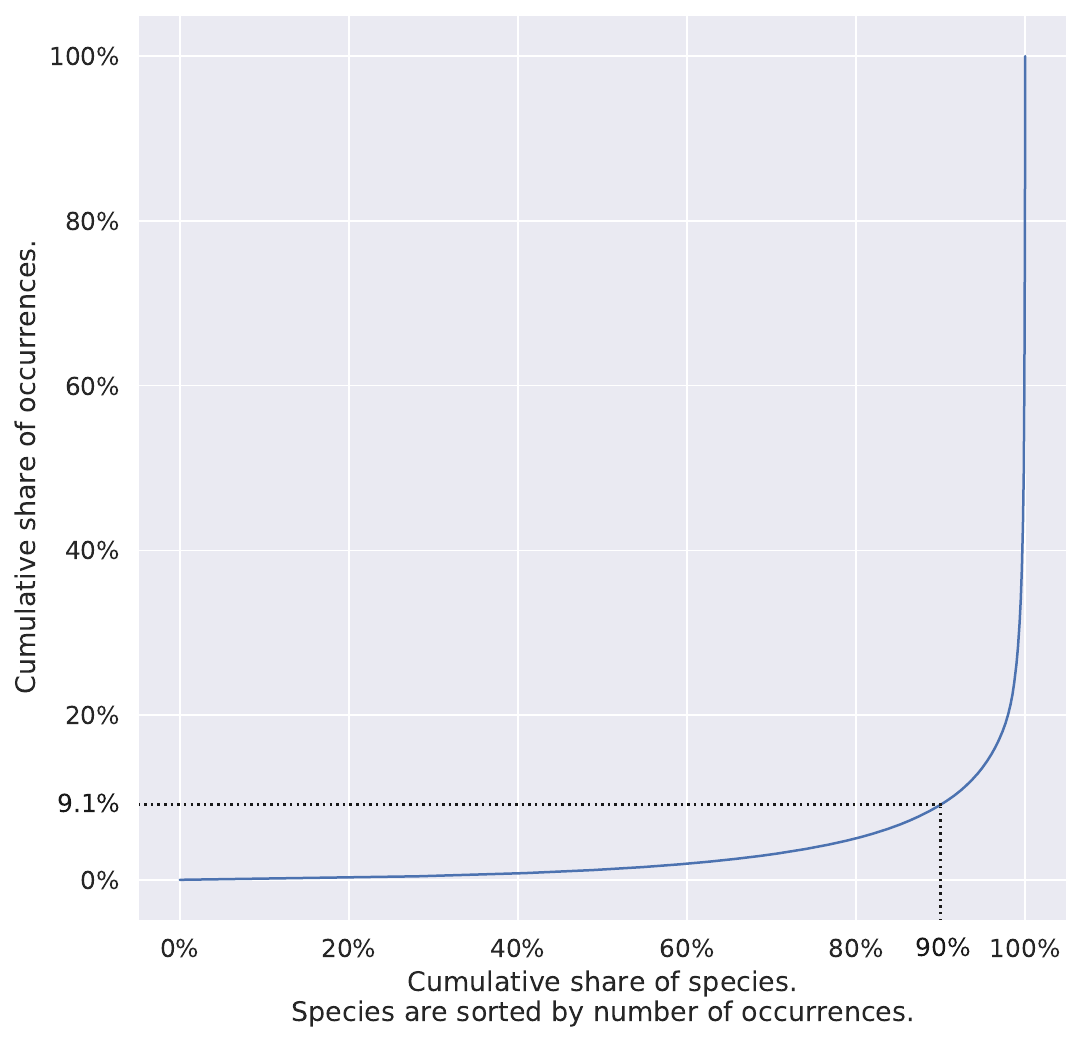}
    \includegraphics[trim={0.2cm 0 0.3cm 0}, clip, width=6.5cm]{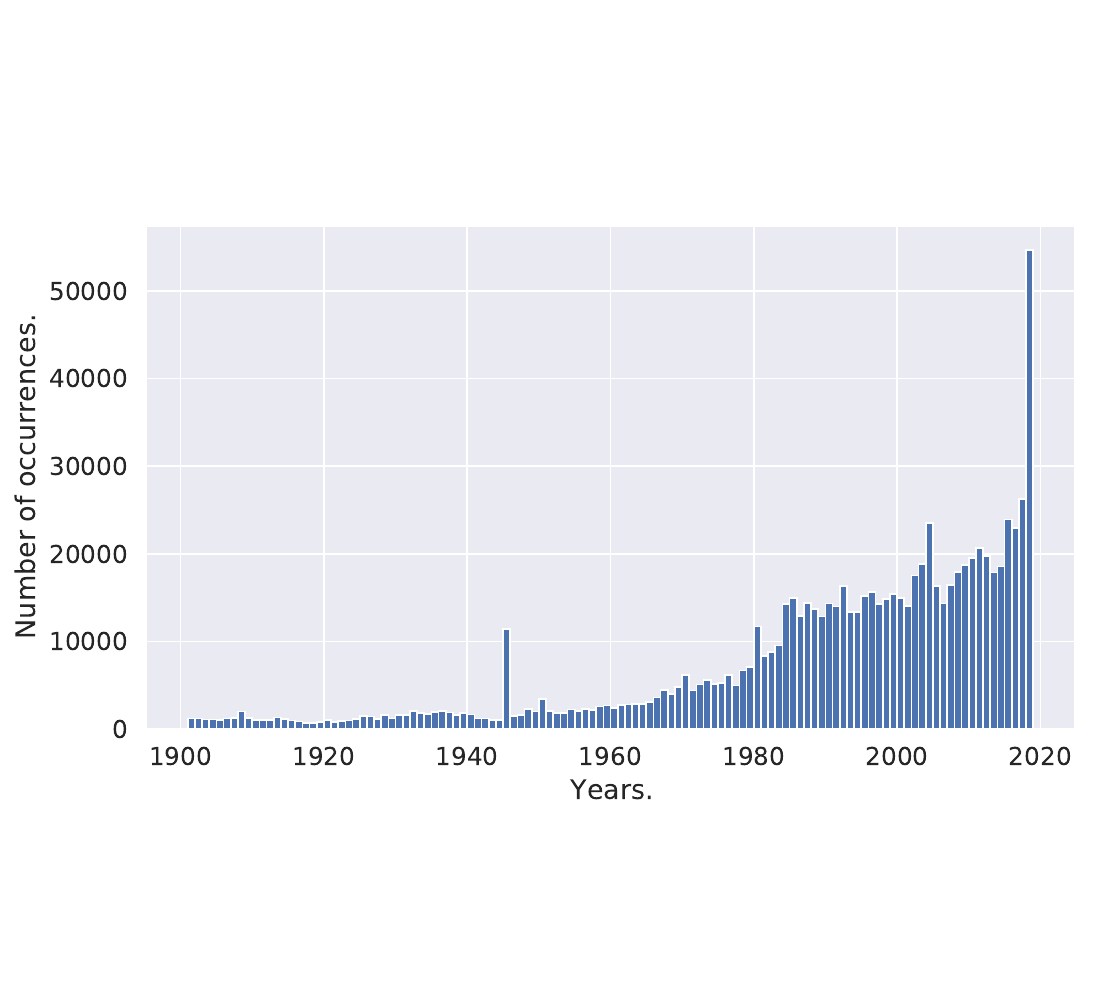}
    \end{center}
    
    \hspace{0.2cm}\textbf{(c)}
    \vspace{-1.15cm}
    \begin{center}
    \includegraphics[width=9.5cm]{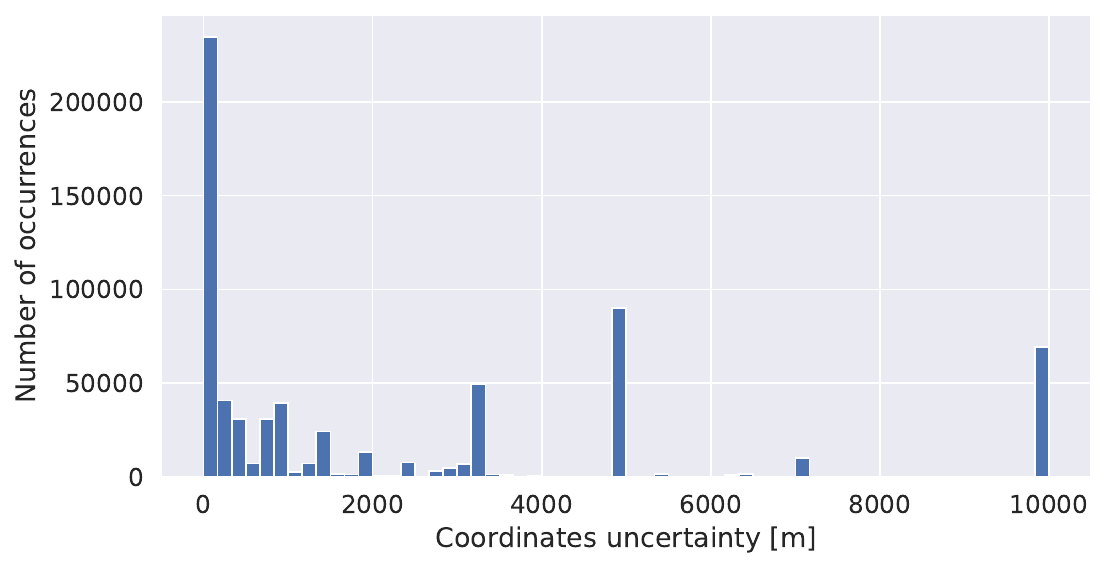}
    \end{center}
    
    \hspace{0.2cm}\textbf{(d)}
    \vspace{-1.15cm}
    \begin{center}
    \includegraphics[trim={3cm 2.5cm 13cm 1.8cm}, clip, width=12.25cm]{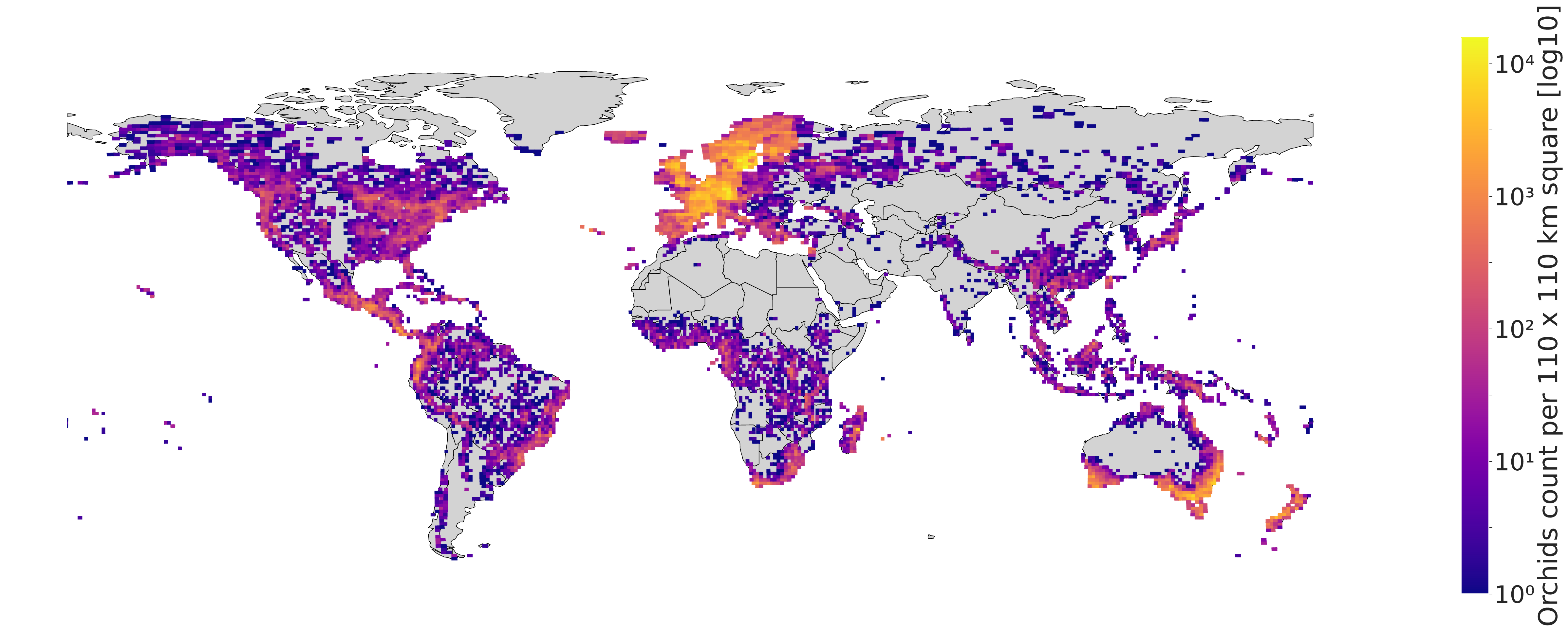}
    \includegraphics[trim={76cm 0 0 0}, clip, width=0.8cm]{Orchids_map_newleg.pdf}
    \end{center}
    
    \vspace{-0.5cm}
    \caption{
    \textbf{(a)} Occurrences' distribution. Species are ordered by frequency. The dotted lines are flagging that 90\% of the species are only gathering 9.1\% of the occurrences.
    \textbf{(b)} Occurrences' temporal distribution. The two previous graphs are based on all dataset's occurrences.
    \textbf{(c)} Histogram of occurrences geolocation uncertainty (60 bins). 31\% of the 999,248 occurrences associated with satellite data had no uncertainty provided at all and are not represented in this figure. Uncertainty was limited to 10,000~m on the Figure. First quartile is 100~m, median is 850~m and third quartile is 5,000~m. Recent and citizen science occurrences are usually integrating quite precise geolocation (explaining left peaks accumulation) whereas old observations will be less precise. The peak at 5,000~m certainly witnesses an arbitrary uncertainty value attributed to part of the orchids.
    While georeferencing uncertainty is effectively a method limitation, deep learning algorithms can statistically learn correct class representations even in the presence of noise \citep{elith_novel_2006, rolnick2017deep}.
    \textbf{(d)} Observations map coloured by number of records in 110 x 110~km tiles (log10 scale).
    }\label{si:distrs_fig}
\end{figure}

\begin{figure}[!htb]
    \begin{center}
        \includegraphics[trim={3cm 1cm 3cm 1cm}, clip, width=\textwidth]{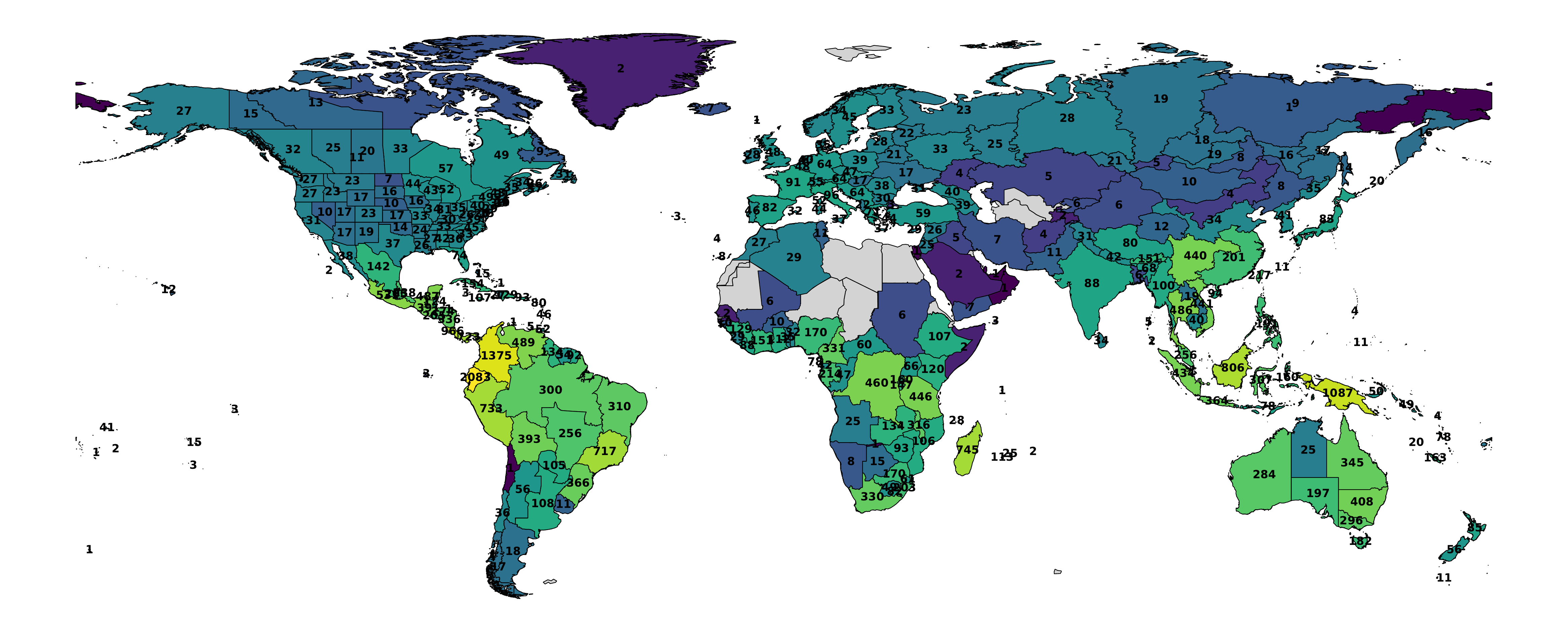}
        \caption{Species richness map stratified by botanical country (WGSRPD level 3). Colours are in log scale.
        }\label{si:richness_map}
    \end{center}
\end{figure}

\section{Predictive features description}
\label{si:predictors}
In selecting predictive features, the main limiting criterion was to use only globally available potential drivers of orchid preferences.

\paragraph{WorldClim2 bioclimatic variables}
The nineteen standard bioclimatic variables from WorldClim version 2 were provided to the model \citep{fick_worldclim_2017}. They are historic averages over the 1970-2000 at 30-second resolution. This suits our occurrence date distribution. Variables stem from temperature and precipitation data (\url{https://www.worldclim.org/data/bioclim.html}). They are established indicators of climate annual trends, seasonality and extreme values.

\paragraph{Soilgrids pedological variables}
Soilgrids is a collection of eleven global soil property and class maps produced by machine learning models \citep{poggio_soilgrids_2021}. They include soil pH, nitrogen concentration annd clay particles proportions among others (more information in the official FAQ: \url{https://www.isric.org/explore/soilgrids/faq-soilgrids}). The exploited statistical models are fitted with 230,000 soil profiles spread wordwide and environmental covariates. We use the 1-kilometre resolution products.

\paragraph{Human footprint detailed rasters}
Eight variables measure direct and indirect global human pressure: built environments, population density, electric infrastructure, crop lands, pasture lands, roads, railways, and navigable waterways \citep{venter_global_2016}. They are provided at a 1-kilometre resolution (\url{https://datadryad.org/stash/dataset/doi:10.5061/dryad.052q5}) and for two distinct years: 1993 and 2009. These rasters spring form both remotely-sensed data and surveys.

\paragraph{Terrestrial ecoregions of the world}
This is a biogeographic classification of terrestrial biodiversity. Ecoregions are defined by the authors as \textit{"relatively large units of land containing a distinct assemblage of natural communities and species, with boundaries that approximate the original extent of natural communities prior to major land-use change"} \citep{olson_terrestrial_2001}. 867 ecoregions are gathered into 14 biomes such as boreal forests or deserts. Data (\url{https://www.worldwildlife.org/publications/terrestrial-ecoregions-of-the-world}) was resampled at 30 seconds longitude/latitude resolution.

\paragraph{Location}
The explicit provision of observation coordinates is a key modelling decision. Both a large regional context and precise location information are provided. The model can make the most of this mixed input. Deep learning models can indeed take advantage of complex combinations of heterogeneous inputs. As longitude and latitude are inputted separately, both indications are processed alongside, can interact, but are also interpreted distinctly.

\begin{figure*}[!htb]
\textbf{(a)}\\
{\centering
\includegraphics[width=\textwidth]{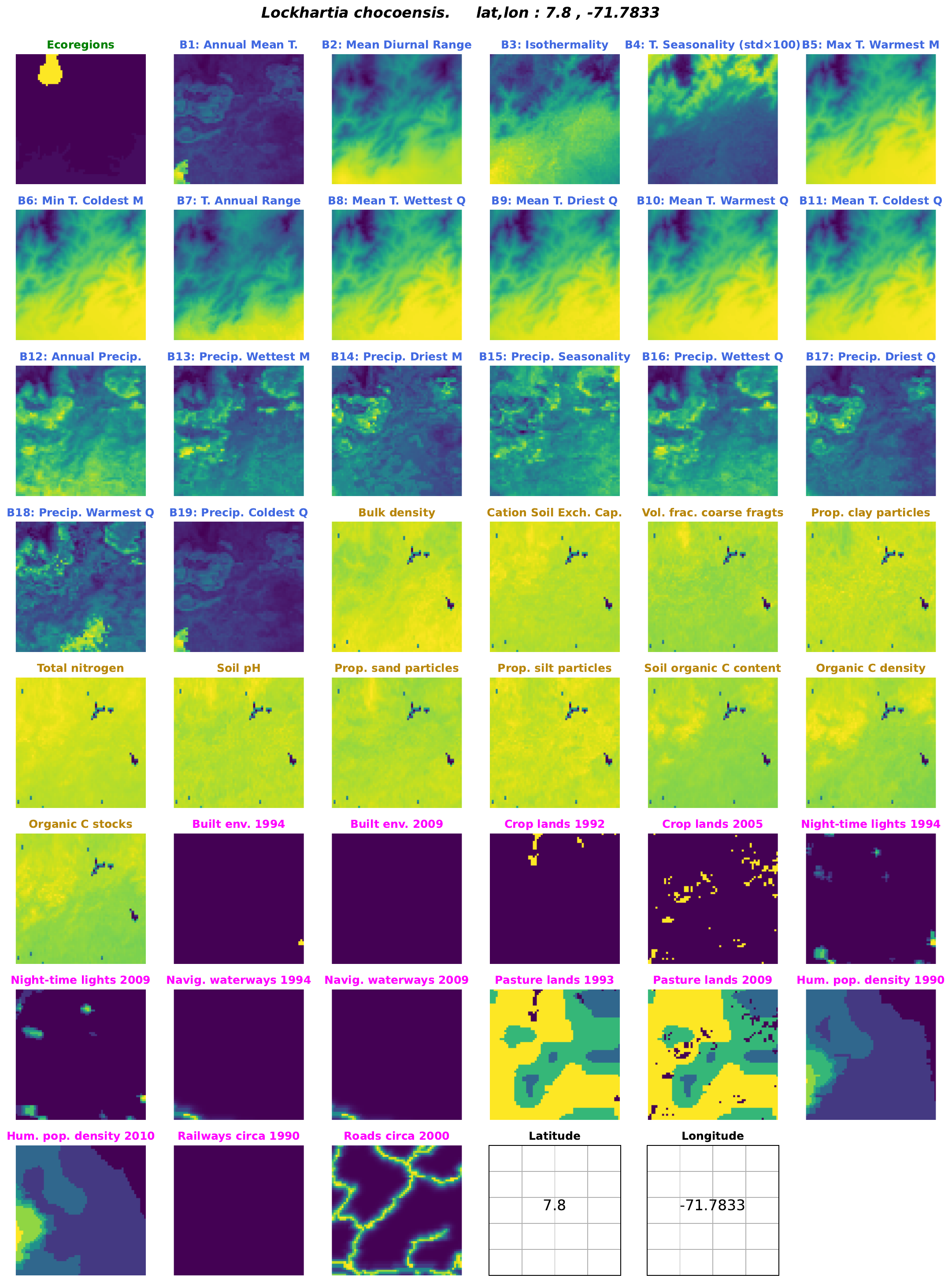}}
\captionsetup{labelformat=empty}
\end{figure*}

\begin{figure*}[!htb]
\textbf{(b)}\\
{\centering
\includegraphics[width=\textwidth]{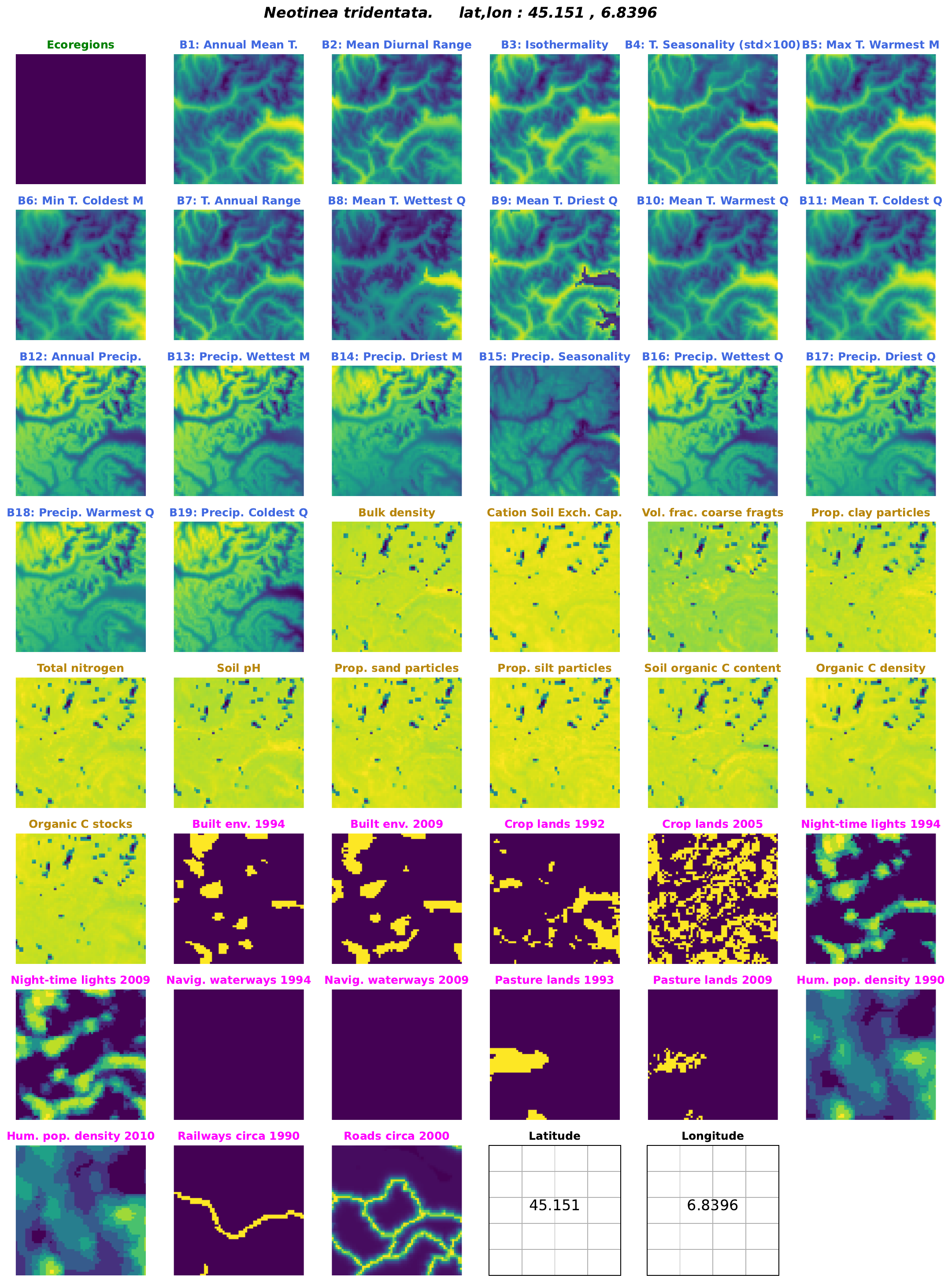}}
\captionsetup{labelformat=empty}
\end{figure*}

\begin{figure}[!htb]
\textbf{(c)}\\
{\centering
\includegraphics[width=\textwidth]{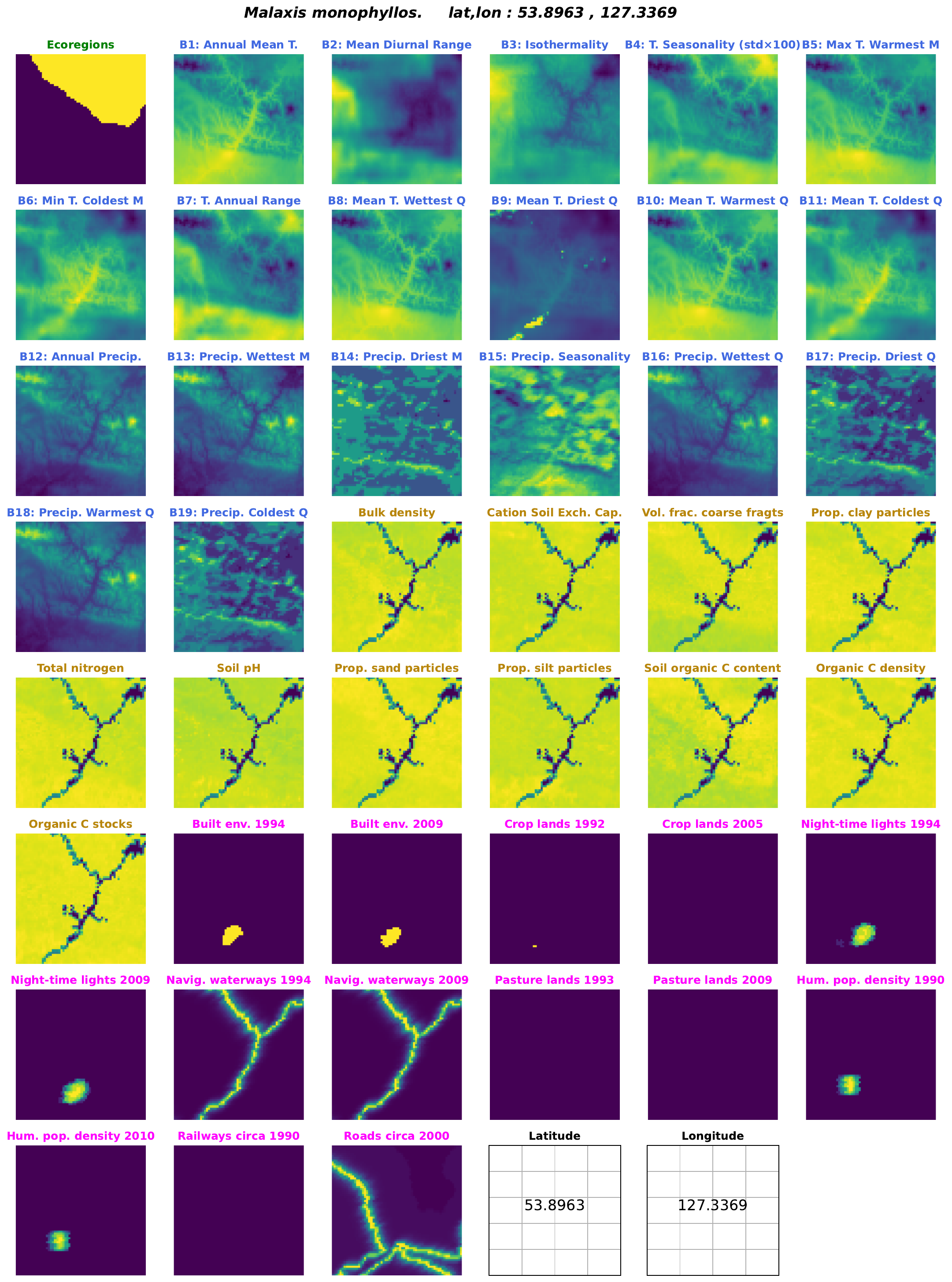}}
\caption{
2D input data associated to three observations located in (a) Venezuela, (b) France and (c) Russia.
Feature types are denoted by their title color:
\textit{(green)} ecoregions, \textit{(blue)} bioclimatic variables, \textit{(brown)} pedological variables, \textit{(pink)} human footprint and \textit{(black)} location.
}\label{si:input_exs}
\end{figure}
\clearpage

\begin{table}[!htb]
\centering
\scriptsize
\begin{threeparttable}
\caption{
List of predictors. They are either categorical or continuous and can be gathered into five groups: the terrestrial ecoregions of the world, the WorldClim2 bioclimatic variables, the Soilgrids pedological variables, the detailed rasters of the human footprint and the location.
}\label{tab:features_table}

\rowcolors{2}{gray!25}{white}
\begin{tabular}{llll}
\headrow
& \textbf{Group} & \textbf{Name} & \textbf{Type}\\
\textbf{1 } &  Terrestrial ecoregions of the world &                               Ecoregions per biome &  categorical \\
\textbf{2 } &     WorldClim2 bioclimatic variables &                     BIO1 = Annual Mean Temperature &   continuous \\
\textbf{3 } &       &  BIO2 = Mean Diurnal Range (Mean of monthly (max temp - min temp)) &   continuous \\
\textbf{4 } &       &            BIO3 = Isothermality (BIO2/BIO7) (×100) &   continuous \\
\textbf{5 } &       &  BIO4 = Temperature Seasonality (standard deviation ×100) &   continuous \\
\textbf{6 } &       &            BIO5 = Max Temperature of Warmest Month &   continuous \\
\textbf{7 } &       &            BIO6 = Min Temperature of Coldest Month &   continuous \\
\textbf{8 } &       &        BIO7 = Temperature Annual Range (BIO5-BIO6) &   continuous \\
\textbf{9 } &       &         BIO8 = Mean Temperature of Wettest Quarter &   continuous \\
\textbf{10} &       &          BIO9 = Mean Temperature of Driest Quarter &   continuous \\
\textbf{11} &       &        BIO10 = Mean Temperature of Warmest Quarter &   continuous \\
\textbf{12} &       &        BIO11 = Mean Temperature of Coldest Quarter &   continuous \\
\textbf{13} &       &                       BIO12 = Annual Precipitation &   continuous \\
\textbf{14} &       &             BIO13 = Precipitation of Wettest Month &   continuous \\
\textbf{15} &       &              BIO14 = Precipitation of Driest Month &   continuous \\
\textbf{16} &       &  BIO15 = Precipitation Seasonality (Coefficient of Variation) &   continuous \\
\textbf{17} &       &           BIO16 = Precipitation of Wettest Quarter &   continuous \\
\textbf{18} &       &            BIO17 = Precipitation of Driest Quarter &   continuous \\
\textbf{19} &       &           BIO18 = Precipitation of Warmest Quarter &   continuous \\
\textbf{20} &       &           BIO19 = Precipitation of Coldest Quarter &   continuous \\
\textbf{21} &      Soilgrids pedological variables &                              Bulk density (cg/cm3) &   continuous \\
\textbf{22} &        &      Cation exchange capacity at ph 7 (mmol(c)/kg) &   continuous \\
\textbf{23} &        &                        Coarse fragments in cm3/dm3 &   continuous \\
\textbf{24} &        &                               Clay content in g/kg &   continuous \\
\textbf{25} &        &                                  Nitrogen in cg/kg &   continuous \\
\textbf{26} &        &                                  pH water (pH *10) &   continuous \\
\textbf{27} &        &                                       Sand in g/kg &   continuous \\
\textbf{28} &        &                                       Silt in g/kg &   continuous \\
\textbf{29} &        &                        Soil organic carbon (dg/kg) &   continuous \\
\textbf{30} &        &                     Organic carbon density (g/dm3) &   continuous \\
\textbf{31} &        &                   Soil organic carbon stock (t/ha) &   continuous \\
\textbf{32} &     Human footprint detailed rasters &  Individual pressure map of built environments in 1994 &  categorical \\
\textbf{33} &       &  Individual pressure map of built environments in 2009 &  categorical \\
\textbf{34} &       &      Individual pressure map of crop lands in 1992 &  categorical \\
\textbf{35} &       &      Individual pressure map of crop lands in 2005 &  categorical \\
\textbf{36} &       &  Individual pressure map of night-time lights in 1994 &  categorical \\
\textbf{37} &       &  Individual pressure map of night-time lights in 2009 &  categorical \\
\textbf{38} &       &  Individual pressure map of navigable waterways in 1994 &   continuous \\
\textbf{39} &       &  Individual pressure map of navigable waterways in 2009 &   continuous \\
\textbf{40} &       &   Individual pressure map of pasture lands in 1993 &  categorical \\
\textbf{41} &       &   Individual pressure map of pasture lands in 2009 &  categorical \\
\textbf{42} &       &  Individual pressure map of human population density in 1990 &  categorical \\
\textbf{43} &       &  Individual pressure map of human population density in 2010 &  categorical \\
\textbf{44} &       &     Individual pressure map of railways circa 1990 &  categorical \\
\textbf{45} &       &        Individual pressure map of roads circa 2000 &   continuous \\
\textbf{46} &                             Location &                                     Longitude (DD) &   continuous \\
\textbf{47} &                              &                                      Latitude (DD) &   continuous \\
\hline
\end{tabular}

\end{threeparttable}
\end{table}

\clearpage

\section{Calibration of the species assemblage prediction model}
\label{si:modelCalib}
The optimisation of the hyper-parameter $\lambda$ is done through an average error control method applied on the validation set. In Equation 3, $\epsilon$ is set to $0.03$. 
The resulting estimated value for $\lambda$ is equal to $8.75\mathrm{e}{-5}$ and the corresponding average size of the predicted species assemblages is equal to $124$ species.
Reaching 0.97\% micro-average accuracy means that the model almost always returns the correct label within the predicted set when a random unseen observation is being provided. The number of observations per class being strongly unbalanced (see Appendix \ref{si:occ_distributions} Fig.~a), the 97\% micro-average accuracy is strongly influenced by the performance on common species.
Now, when all unseen species are granted the same weight in the average computation (macro-average accuracy), performance is still of 80\%. Given how unbalanced the observation dataset is (median occurrence number is four, 25\% species have more than 13 occurrences), it becomes clearer that the model's performances are satisfying.
Summary statistics on $|\hat{S}_\lambda|$ are reported on Table \ref{si:Tstats}.\\

\begin{figure}[H]
    \begin{center}
        \frame{\includegraphics[width=13cm]{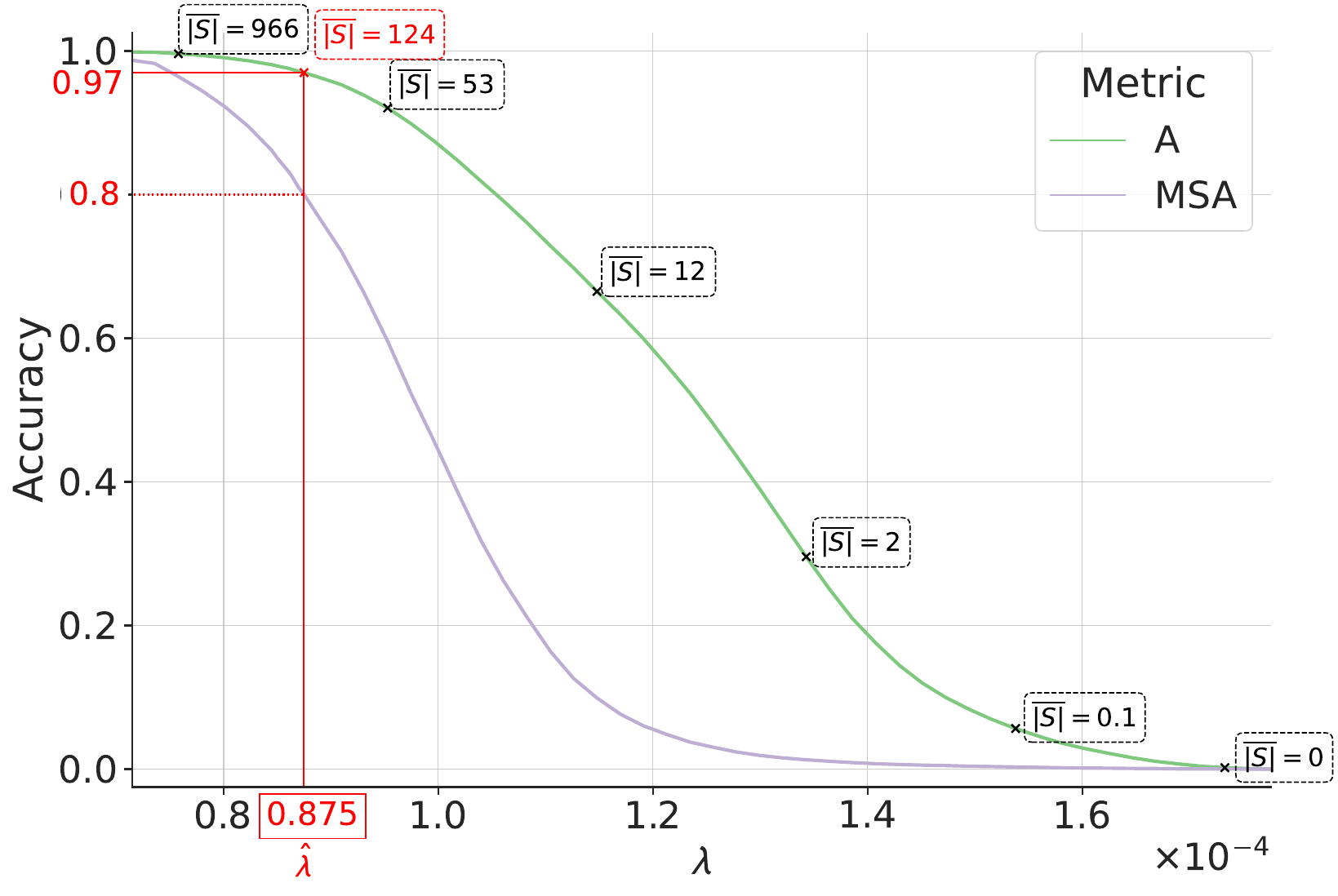}}
        \caption{ 
        Average error control setting on the validation set. Limiting condition on the micro-average accuracy (green curve) is  $\epsilon \leq 0.03 \Leftrightarrow \text{A} \geq 0.97$.  Optimal threshold $\hat{\lambda}$ is highlighted in red while matching macro-average accuracy (grey function) is also reported with a red dashed line. Average set sizes $\overline{|\hat{S}_\lambda|}$ are indicated in dashed boxes (hat and subscript being dropped for readability).
        }\label{si:model_calib_fig}
    \end{center}
\end{figure}

\begin{table}[H]
    \begin{center}
        \caption{ 
        Validation set statistics on $|\hat{S}_\lambda|$, i.e. the size of the species assemblage after thresholding the conditional probabilities of presence with $\hat{\lambda}$ (46,290 validation points).
        The minimum number of species retained in the validation set is four.
        However, on a global scale there are areas with no species above $\hat{\lambda}$, resulting in empty predictions (e.g. western Algeria).
        It is also very likely that areas are predicted with more than 401 orchids (the maximum on the validation set).
        }\label{si:Tstats}
        \begin{threeparttable}
            \rowcolors{2}{gray!25}{white}
            \begin{tabular}{lrrrrrrrrr}
                \headrow
                {} & \textbf{mean} & \textbf{std} & \textbf{min} & \textbf{25\%} & \textbf{50\%} & \textbf{75\%} & \textbf{max} & \textbf{A}$_{30}$ & \textbf{MSA}$_{30}$\\
                \textbf{$\hat{\lambda}$} &  124.067 &  39.803 &    4 &   95 &  121 &  150 &  401 &               0.970 &                0.801 \\
                \hline
            \end{tabular}
        \end{threeparttable}
    \end{center}
\end{table}
\clearpage

\section{Confusion matrices and relative change in the number of threatened species}
\vspace{-0.5cm}
\begin{figure}[H]
    \centering
    \begin{minipage}{.5\textwidth}
        \centering
        \textbf{(a)}\\
        \includegraphics[width=\linewidth]{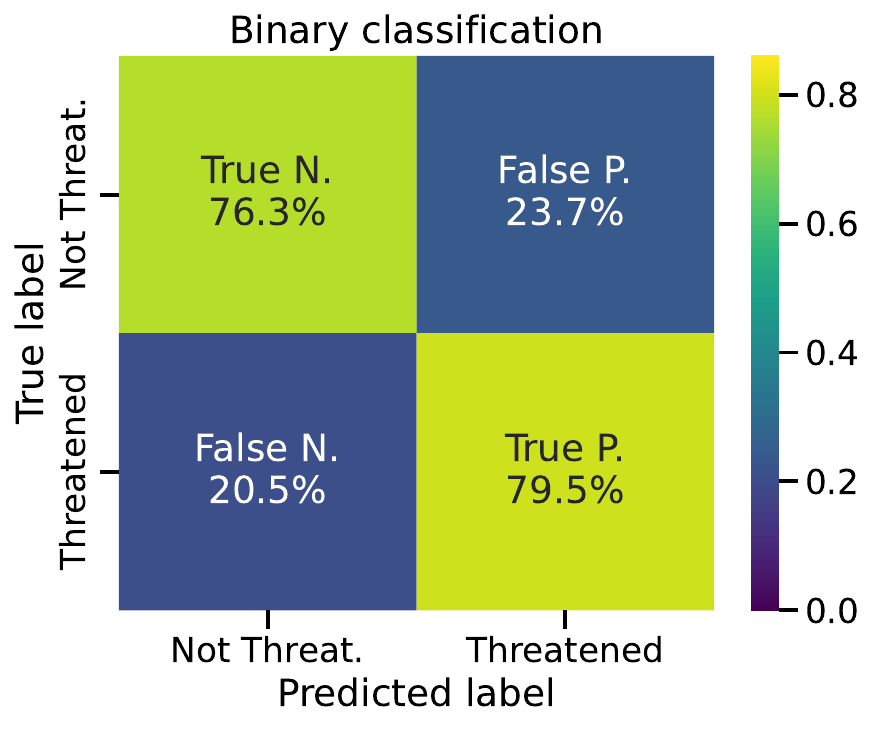}
    \end{minipage}%
    \begin{minipage}{0.5\textwidth}
        \centering
        \textbf{(b)}\\
        \includegraphics[width=\linewidth]{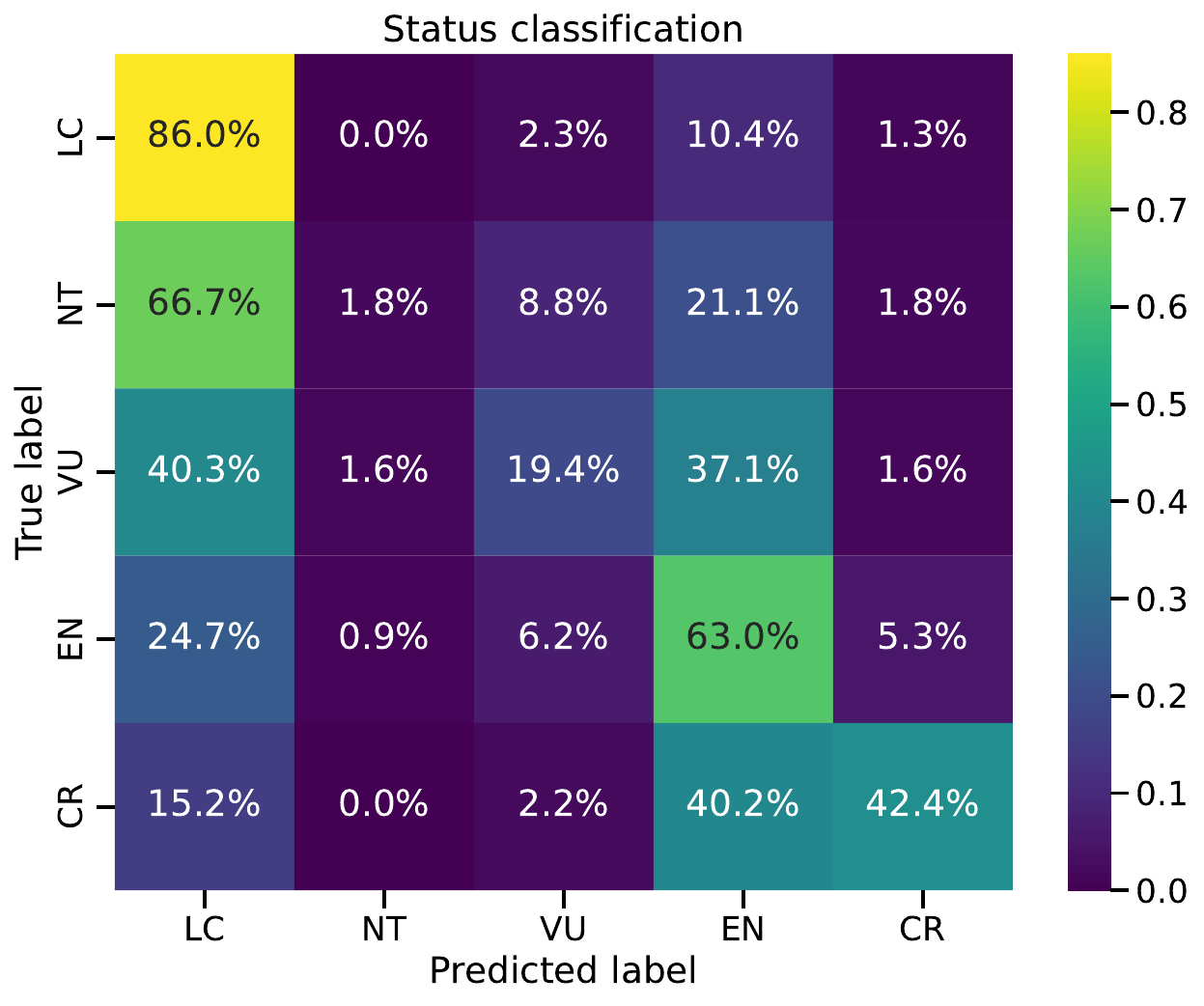}
    \end{minipage}
    \begin{minipage}{.5\textwidth}
        \centering
        \textbf{(c)}\\
        \includegraphics[width=\linewidth]{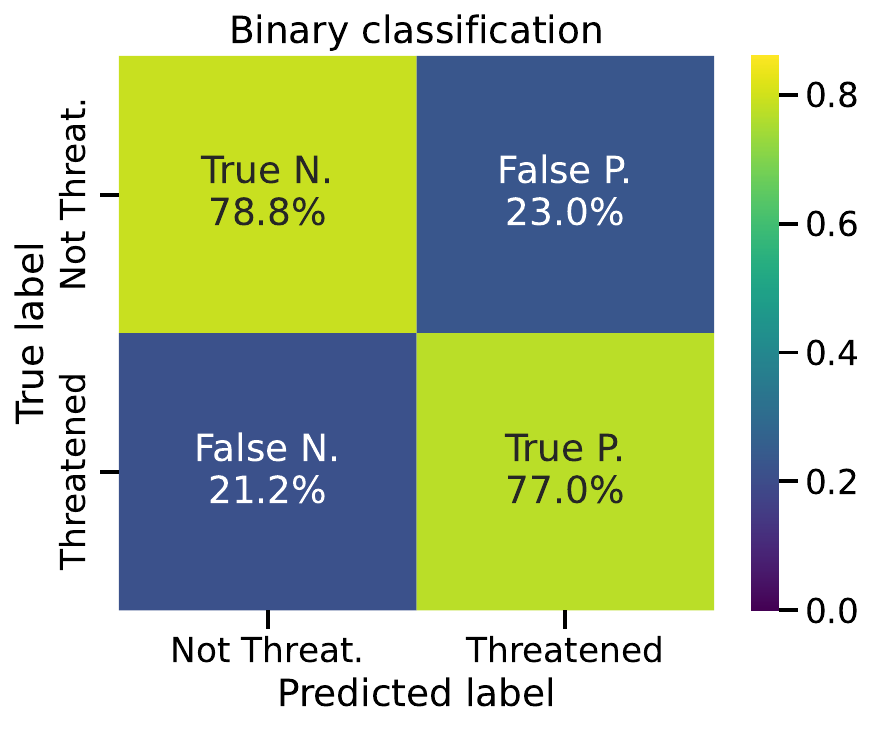}
    \end{minipage}%
    \begin{minipage}{0.5\textwidth}
        \centering
        \textbf{(d)}\\
        \includegraphics[width=\linewidth]{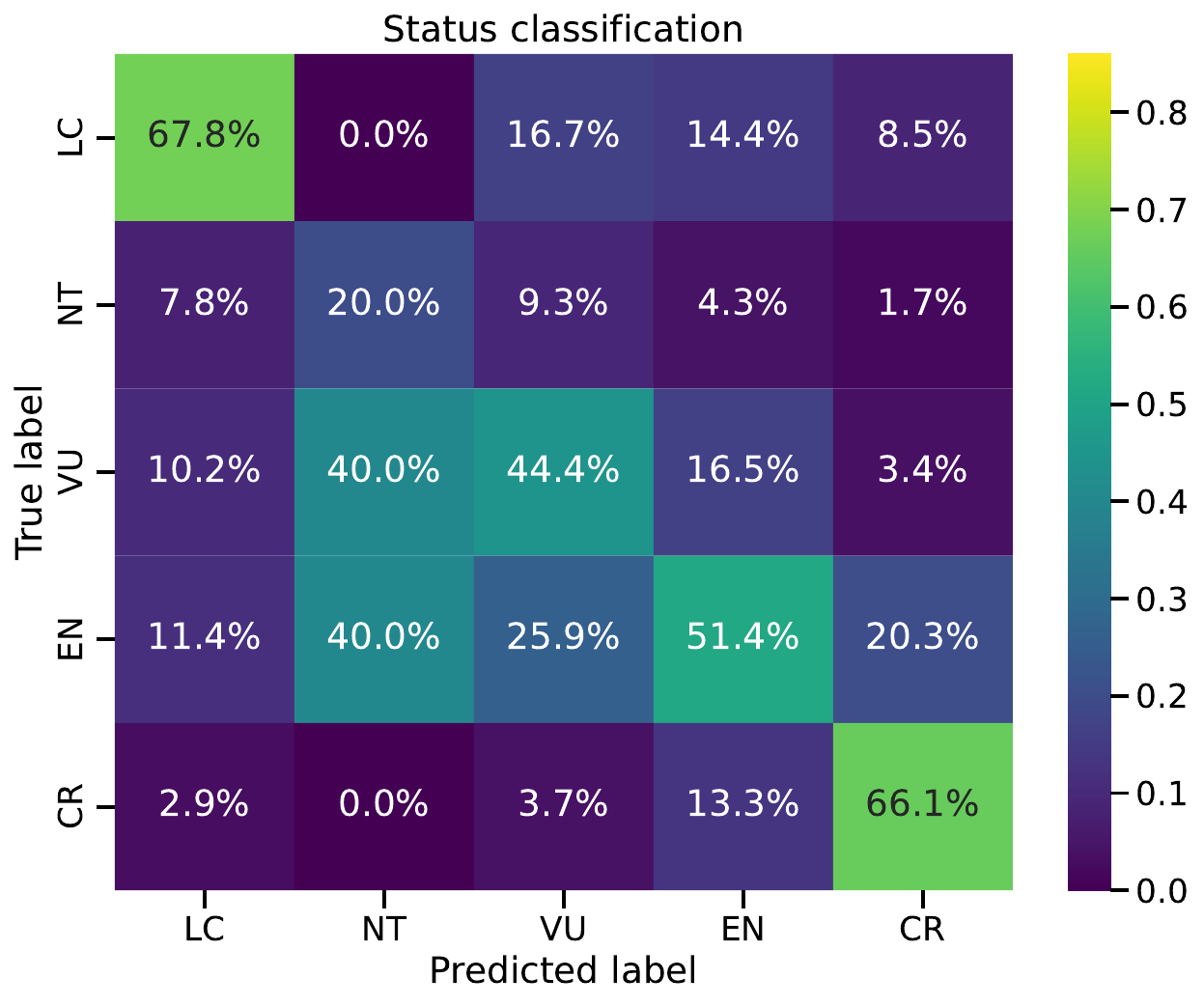}
    \end{minipage}
    \caption{
    The first row shows the precision confusion matrices, i.e. with row normalisation. Among the true labels of a given category, we test the proportions predicted in each category.
    The second row presents recall confusion matrices, i.e. with a normalisation by column. Among the predicted labels of a given category, we test the proportions that actually belong to the predicted category or to others.
    The first column shows the results for the binary classification and the second column for the status classification.\\
    \textbf{(a)} Precision confusion matrix for binary classification.
    79.5\% of threatened species are correctly classified as threatened by the model.\\
    \textbf{(b)} Precision confusion matrix for status classification.
    86\% of LC species are correctly identified by the model.
    For NT species, 66.7\% are misclassified as LC and 21.1\% as EN.
    40.3\% of the VU species are misclassified as LC, 37.1\% as EN and 19.4\% are correctly classified.
    Three quarters of the EN species are classified as either VU, EN or CR.
    Of the species classified as CR by IUCN, the model predicts 40.2\% as EN and 42.4\% actually as CR.\\
    \textbf{(c)} Recall confusion matrix for binary classification.
    78.8\% of species predicted as not threatened are actually so.\\
    \textbf{(d)} Recall confusion matrix for status classification.
    Only 67.8\% of species predicted as LC are actually Least Concern.
    Of those predicted to be NT, only a fifth are actually NT, with the remainder split evenly between VU and EN.
    44.4\% of species predicted to be VU are correctly classified.
    More than half of the species predicted as EN are actually EN, and more than 80\% are from a threatened category.
    Of the species predicted as CR, 86.4\% are actually classified as either CR or EN by IUCN.
    }
    \label{si:cf_mat}
\end{figure}

\begin{figure}[!htb]
\begin{center}
    \includegraphics[width=0.6\textwidth]{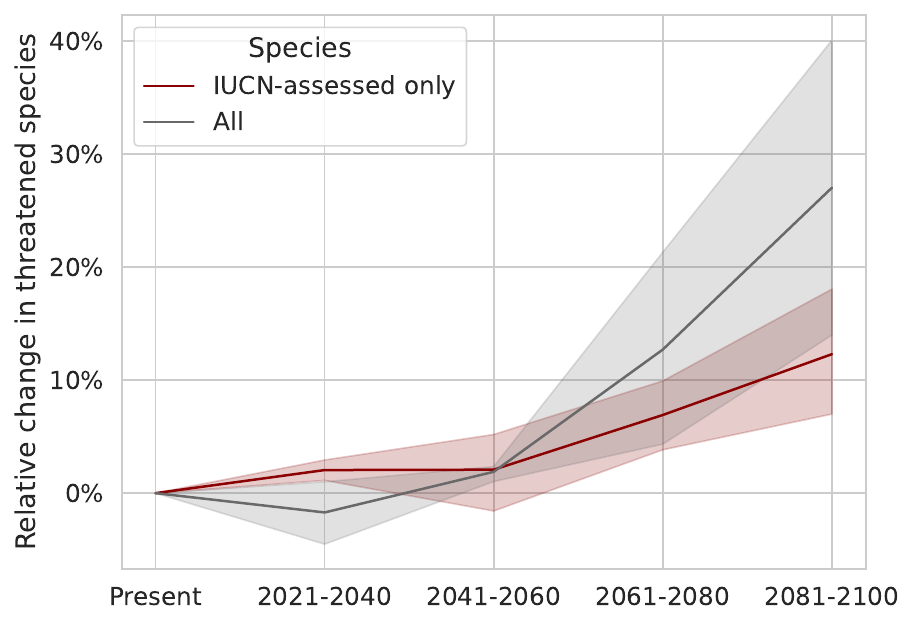}
\end{center}
\caption{Relative change in the number of species predicted to be threatened over time.
The bands indicate the uncertainty due to the combination of the two dispersal scenarios.
Interestingly, for species not yet assessed by the IUCN, the increase in threatened species is predicted to be higher by the end of the century.
}\label{si:global_threat}
\end{figure}
\clearpage

\section{Description of the \textit{.csv} files containing the status predictions}
\label{si:status_pred}
\paragraph{Description of the status prediction supplementary file \textit{ALL\_species\_status.csv}.}
\noindent This \textit{.csv} lists the status predicted with our classifier for the 14,129 species of our dataset:
\begin{itemize}[nosep]
    \item[-] At two different levels: \textit{broad}, i.e. the binary classification Threatened or not, and \textit{detail}, i.e. the five IUCN categories from LC to CR.
    \item [-] For two different dispersal scenario: \textit{False} meaning that no dispersal is allowed and species can only be re-predicted only where they once occurred, and \textit{True} meaning that species can potentially be re-predicted on all our dataset points.
    \item[-] For five different time periods matching the Worldclim 2 bioclimatic projections: \textit{Present, 2021-2040, 2041-2060, 2061-2080} and \textit{2081-2100}.\\
\end{itemize}

Here are the meanings of the common fields:
\begin{itemize}[nosep]
    \item[-] \textit{species} is the GBIF canonical species name
    \item [-] \textit{speciesKey} is the GBIF unique key associated to the species
    \item [-] \textit{lat, lon, HFP09, Elevation} are the species average values across the dataset's observations for respectively \textit{latitude, longitude, 2009 human footprint index} and \textit{elevation}.
    \item[-] \textit{continents} provides the set of WGSRPD level 3 regions in which the species occurs.
    \item[-] \textit{IUCNonly} indicates whether the species is currently assessed by the IUCN (886) or not (13,243).
    \item[-] For IUCN-assessed species, the present categories are the current official red list levels and not predictions.\\
\end{itemize}

\paragraph{Description of the Extinct species supplementary file \textit{EXT\_species.csv}.}
\noindent This \textit{.csv} lists the 234 species predicted to be extinct by the model.
Fields are identical to the \textit{ALL\_species\_status.csv} file described below, with the noticeable differences being:
\begin{itemize}[nosep]
    \item[-] \textit{time period} is the field indicating \underline{the first period} from which the species is predicted to be extinct.
    \item[-] All species predicted to be extinct come from the null dispersal scenario.
\end{itemize}
In addition, one species is predicted to be extinct as of now: \textit{Dendrobium rhytidothece}. This species is not red listed and is known on the GBIF only from six occurrences, five of which date from 1909 and the \href{https://www.gbif.org/species/5315531}{last one} being fuzzy.
There are 42 species currently assessed by the IUCN that are predicted to become extinct (and 192 species that have not yet been assessed).
\clearpage

\section{Results considering only IUCN-assessed species}
\label{si:iucn-assessed_only}
\begin{figure}[!htb]
{\centering
\includegraphics[width=\textwidth]{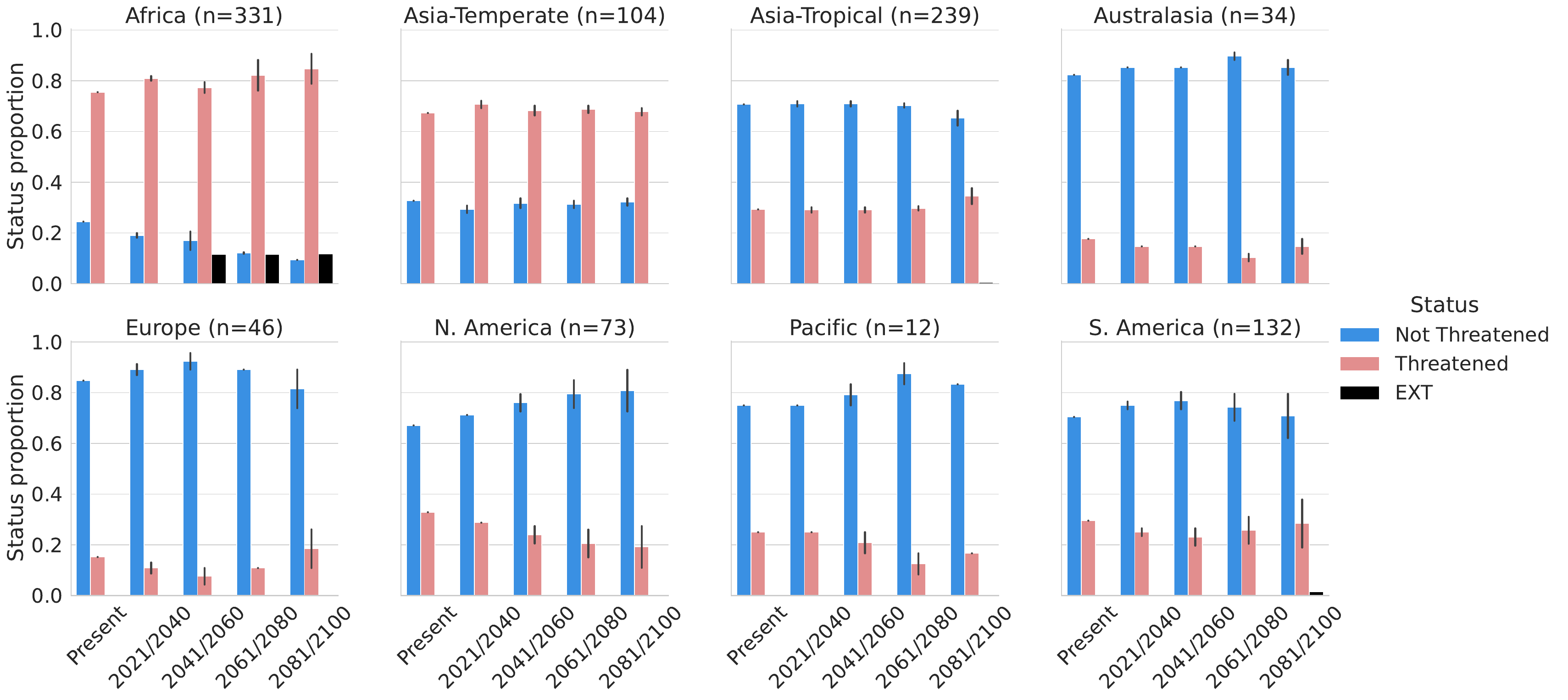}
}
\caption{Binary status proportions per continent over time periods. Only IUCN-listed species are included, and their number per continent is given in the subtitle. Error bars account for differences between the two dispersal scenarios.
With this restriction, some different trends are apparent.
In North America, out of 73 species, the proportion of threatened species is actually predicted to decrease steadily (same trend for the 12 Pacific species, but there are too few species to be considered a robust result).
In Africa, the proportion of threatened species is projected to increase to more than 80\% by the end of the century. 
In tropical Asia, the number of threatened species is increasing but is significantly lower than when all species are considered.
Levels are also lower in Europe if only IUCN-assessed species are considered.
Overall, it is interesting to observe that continental trends and levels can be quite different when only IUCN-assessed species are considered.
For us, this is an indication of the model's ability to generalise without simply overfitting and replicating patterns between neighbouring species.
}\label{fig:conts_iucn}
\end{figure}

\begin{figure}[H]
{\centering
\includegraphics[width=\textwidth]{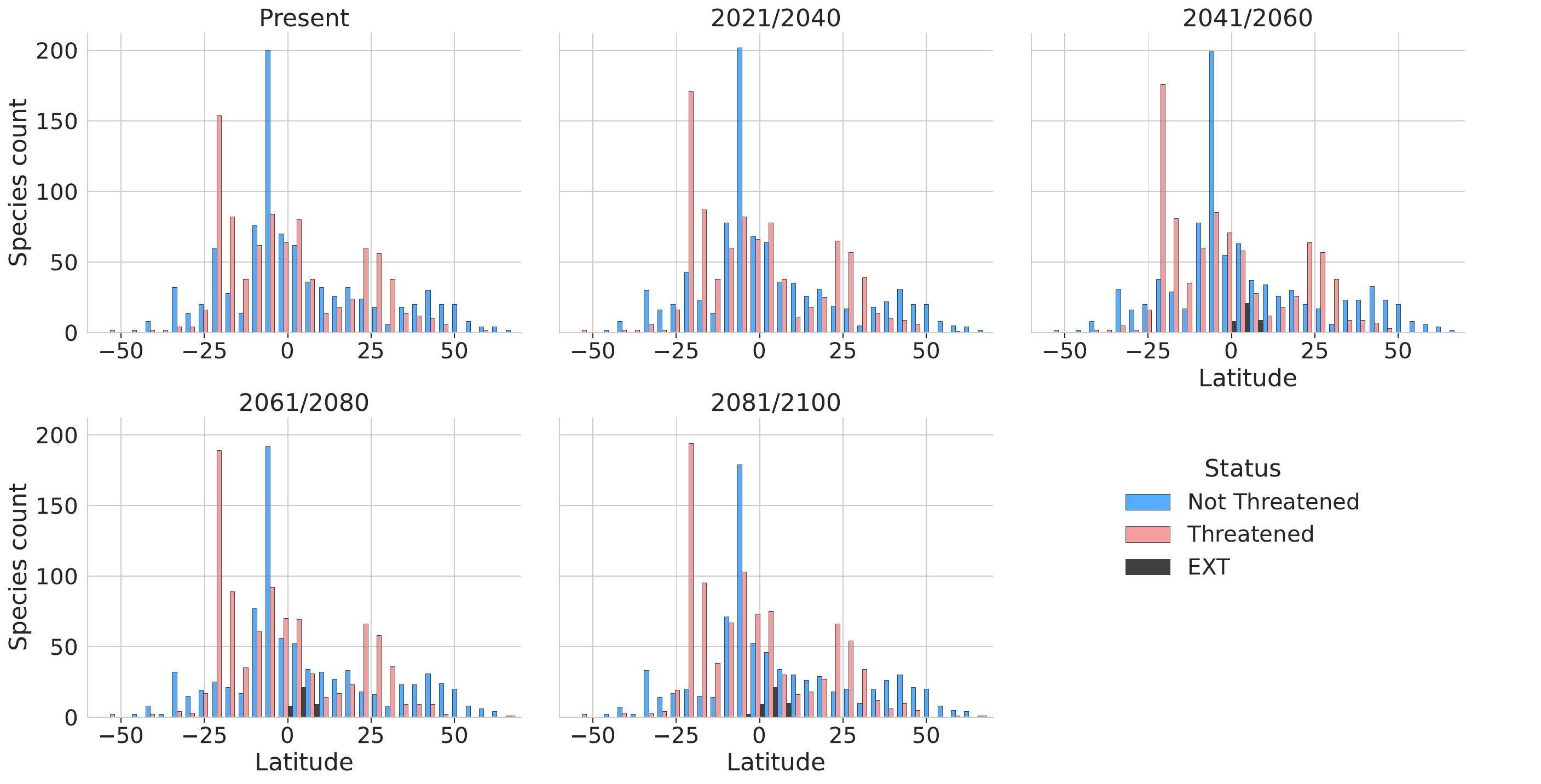}
}
\caption{Species count histograms as a function of average latitude and over time. Only IUCN-listed species are included. The bins cover four degrees of latitude.
A trident pattern appears for threatened species, but: i) the highest threatened species peak is at about -20° latitude and ii) high counts of threatened species around the 25° parallel do not increase, unlike the figure for all species.
Again, the differences between the two species support confirms that our status classifier relies on specific species-level information and not just spatial information.
This figure also confirms that the current IUCN assessment is biased towards the northern hemisphere.
}\label{fig:lat_iucn}
\end{figure}

\begin{figure}[H]
{\centering
\includegraphics[width=\textwidth]{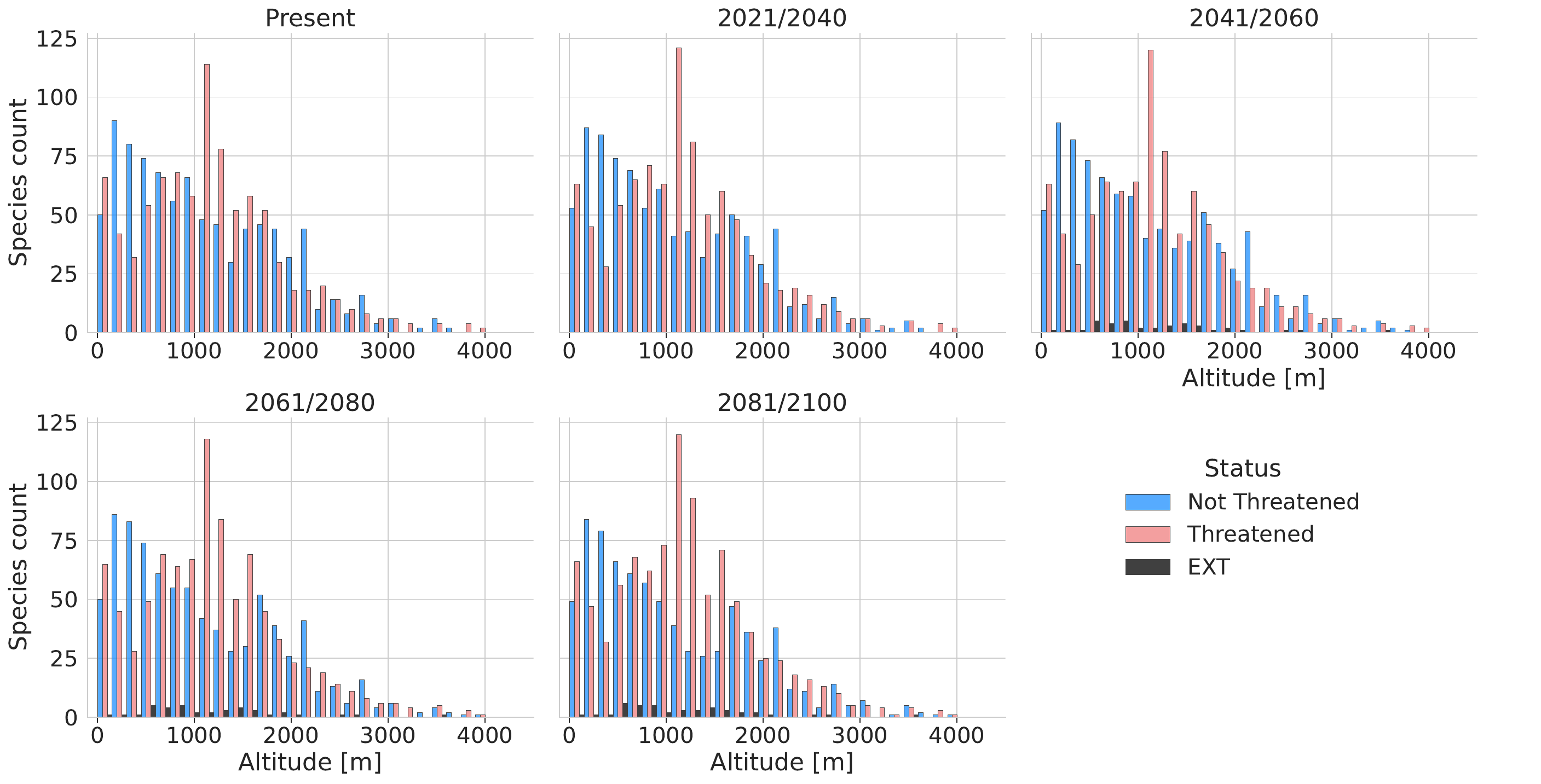}
}
\caption{Species count histograms as a function of average altitude and over time. Only IUCN-listed species are included. Boxes cover 150 metre altitude ranges.
Overall, we found the same patterns in this figure as in its all-species counterpart.
However, the hump-shaped concentration of threatened species at 800-1,500m is replaced here by a more irregular peak forest.
Therefore, in this case, our approach seems to regularise the number of threatened species along the altitudinal gradient.
Nevertheless, causality cannot be inferred from our study.
A confounding variable such as threat exposure could indeed be at the origin of this pattern.
In addition, species already IUCN-assessed and predicted to become extinct seem to be distributed between the lowlands and around 2000~m.
}\label{fig:elev_iucn}
\end{figure}
\clearpage

\section{Status in function of human footprint}
\label{si:hfp_section}
\begin{figure}[!htb]
\begin{center}
    {\includegraphics[width=0.75\textwidth]{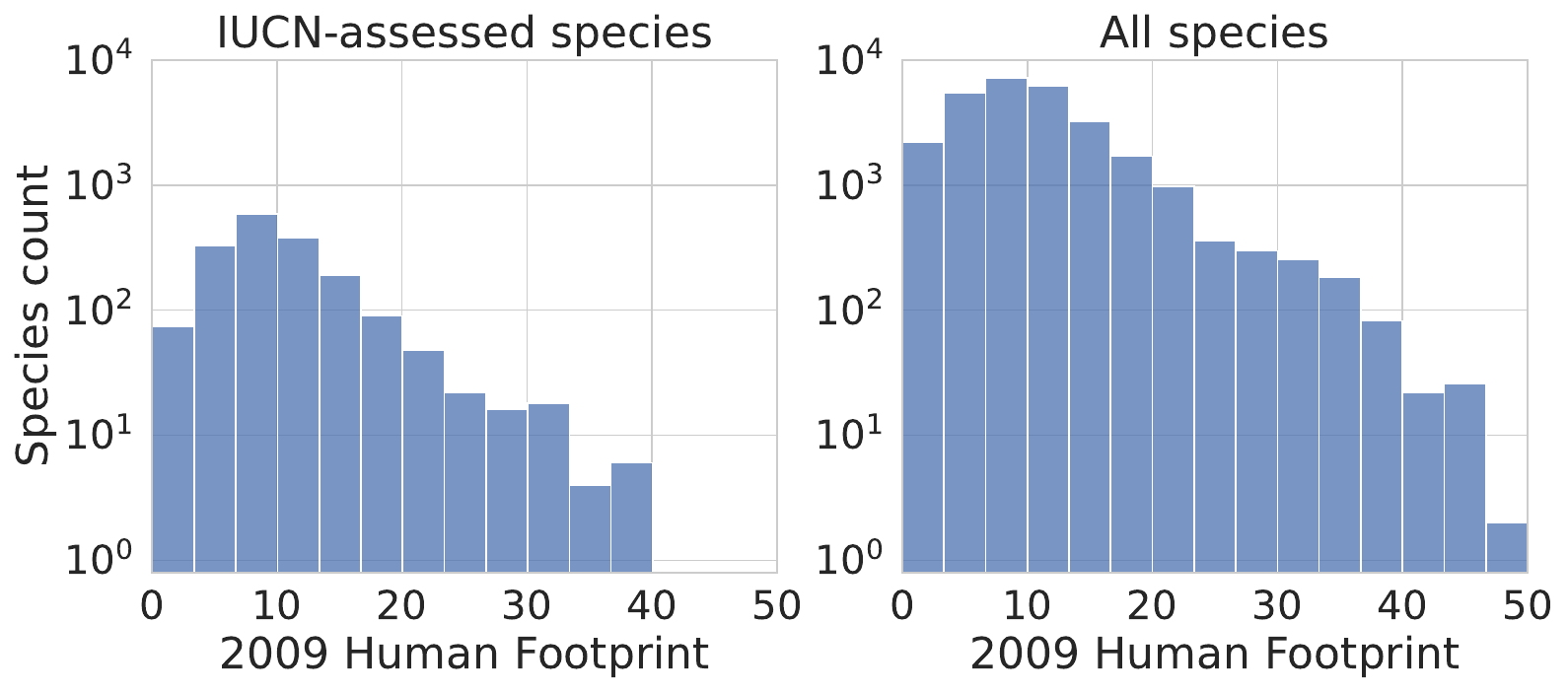}}
\end{center}
\caption{Histogram of the number of species per 2009 human footprint (HFP) bin, considering: \textit{(left)} IUCN-assessed species only and \textit{(right)} all species \citep{venter_global_2016}.
The species count axis is \textbf{log-scale}.
The HFP score of a species is averaged over its current occurrences.
These histograms allow us to assess the number of species per bin used to calculate the proportions shown in Figure \ref{fig:HFP}.
The main message is that high HFP bins contain very few species and the following proportions should then be treated with caution.}\label{fig:hists}
\end{figure}
\clearpage

\begin{figure}[!htb]
\textbf{(a)}\\
{\centering \includegraphics[width=\textwidth]{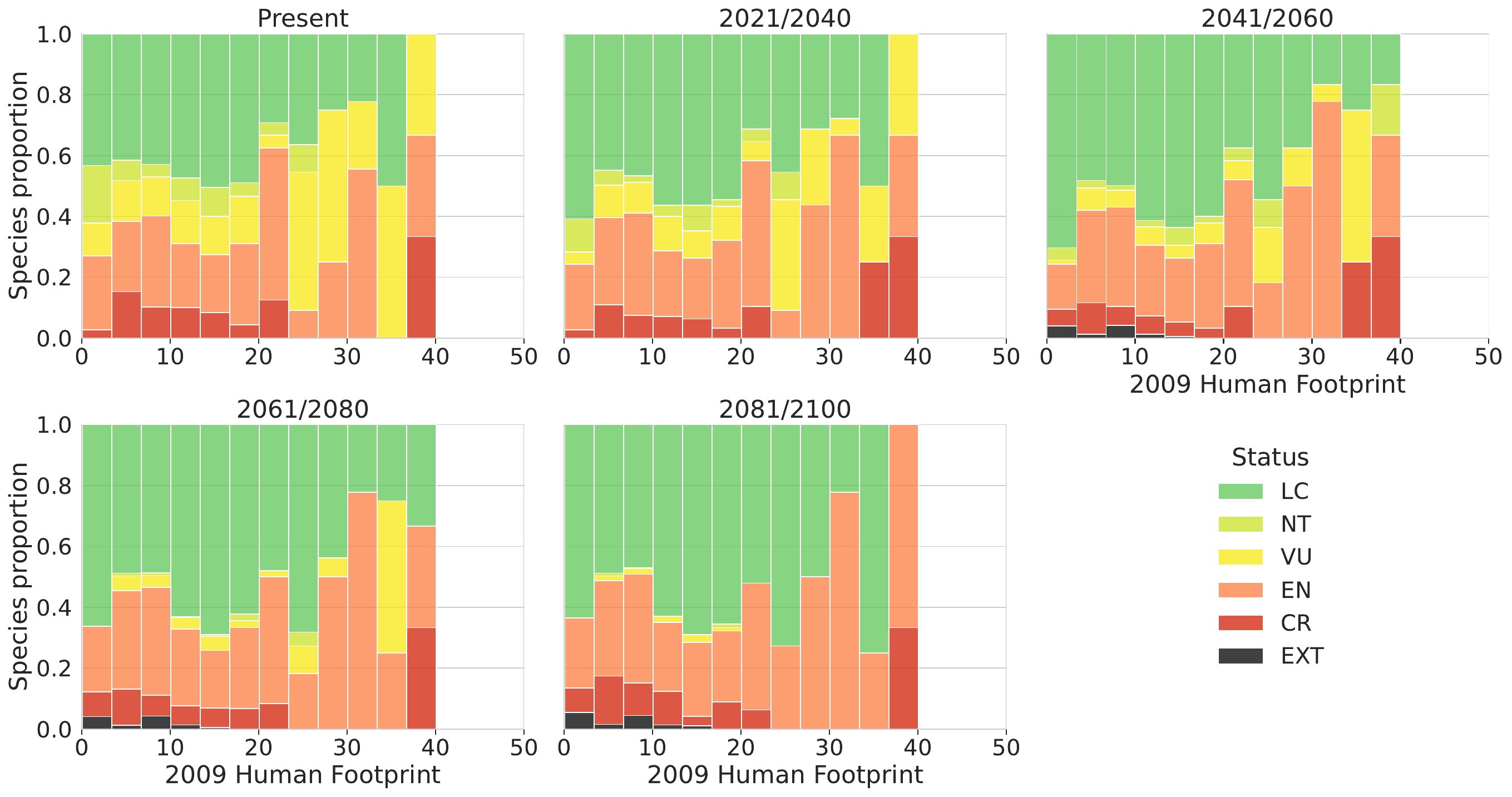}}
\textbf{(b)}\\
{\centering \includegraphics[width=\textwidth]{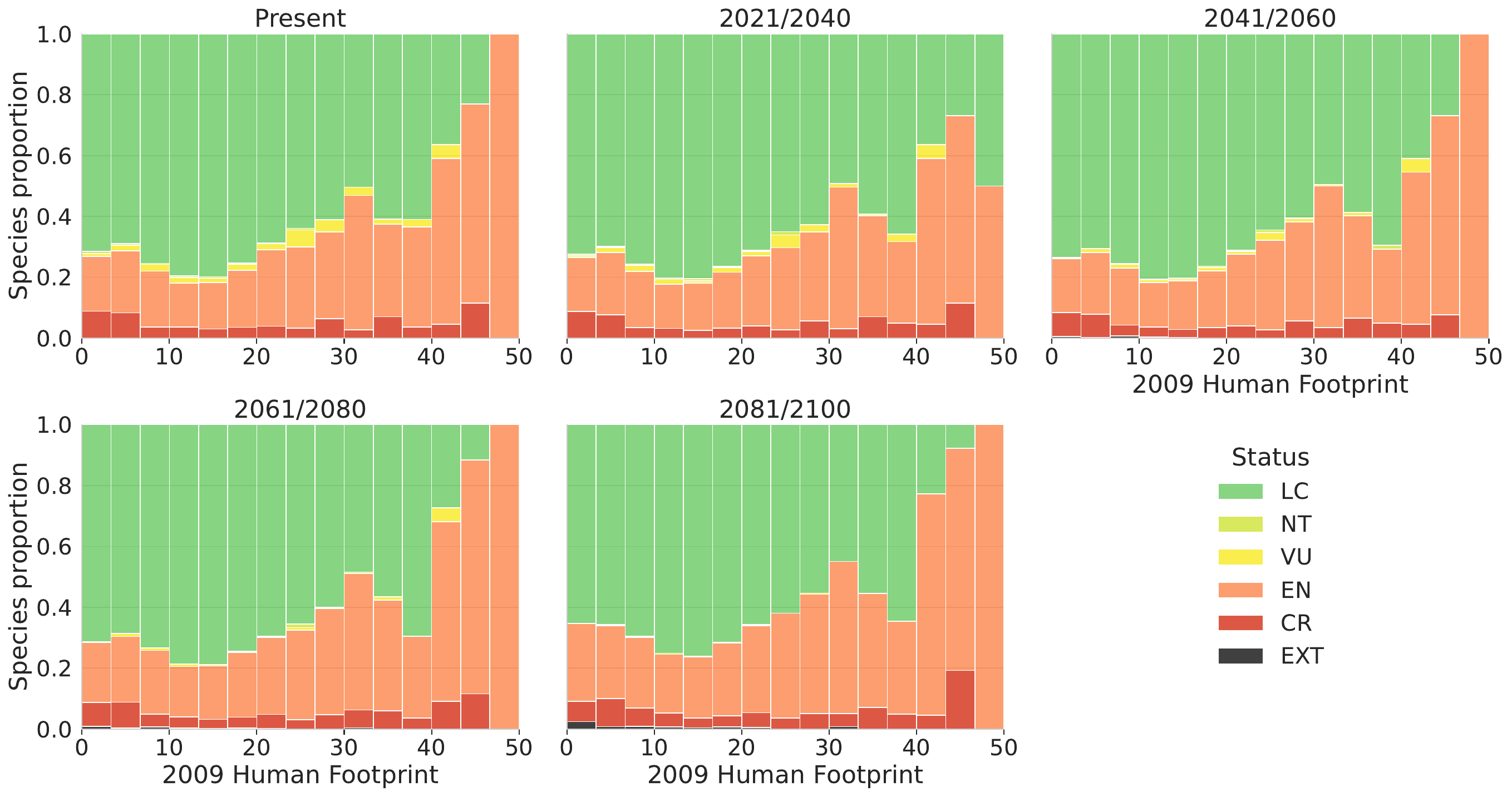}}
\caption{
Predicted status proportions per 2009 human footprint (HFP) bin across time periods, considering \textbf{(a)} IUCN-assessed species only or \textbf{(b)} all species in our dataset.
\textbf{(a)} \textit{Present} sub-figure then represents the HFP distribution of currently red listed orchids.
We observe that: i) NT and VU proportions are currently significant according to IUCN, but are predicted to disappear over time.
This is due to poor classifier performance on these categories and their rather confused definitions, see status confusion matrices Figure \ref{si:cf_mat} and the Discussion.
ii) In both cases, the proportions of threatened species appear to be positively correlated with HFP intervals.
iii) HFP bins may cover few species (see Figure \ref{fig:hists}), so the robustness of these results should be further assessed.
iv) As with the altitude study, this result shows a correlation, but no causality can be inferred. Again, a confounding variable may be at the origin of this pattern rather than HFP.
}\label{fig:HFP}
\end{figure}

\end{document}